\documentclass{ieeeaccess}
\usepackage{cite}
\usepackage{amsmath,amssymb,amsfonts}
\usepackage{algorithmic}
\usepackage{textgreek}
\usepackage{graphicx}
\usepackage{textcomp}
\usepackage[version=4]{mhchem}
\usepackage{hyperref}
\usepackage{braket}
\usepackage{caption}

\def\BibTeX{{\rm B\kern-.05em{\sc i\kern-.025em b}\kern-.08em
    T\kern-.1667em\lower.7ex\hbox{E}\kern-.125emX}}
\begin{document}
\history{Date submitted 24 February 2026}
\doi{Not yet issued}

\title{Toward a CMOS-integrated quantum diamond biosensor based on NV centers}
\author{
    \uppercase{Ioannis Varveris}\authorrefmark{1,2}, 
    \uppercase{Gianni D. Aliberti}\authorrefmark{1,2},
    \uppercase{Felix J. Barzilaij}\authorrefmark{2},
    \uppercase{Zhi Jin}\authorrefmark{2},
    \uppercase{Samantha A. van Rijs}\authorrefmark{2},
    \uppercase{Qiangrui Dong}\authorrefmark{3},
    \uppercase{Daan Brinks}\authorrefmark{3},
    \uppercase{Salahuddin Nur}\authorrefmark{2}, and
    \uppercase{Ryoichi Ishihara}\authorrefmark{1,2}
}

\address[1]{QuTech, Delft University of Technology, Lorentzweg 1, 2628 CJ, Delft, The Netherlands}
\address[2]{Department of Quantum \& Computer Engineering, Delft University of Technology, Mekelweg 4, 2628 CD Delft, The Netherlands}
\address[3]{Imaging Physics Department, Delft University of Technology, Lorentzweg 1, 2628 CJ, Delft, The Netherlands}
\tfootnote{}

\markboth
{Varveris \headeretal: Toward a CMOS-integrated quantum diamond biosensor}
{Varveris \headeretal: Toward a CMOS-integrated quantum diamond biosensor}

\corresp{Corresponding author: Gianni D. Aliberti (email: g.d.aliberti@tudelft.nl)}

\begin{abstract}
We report progress toward a CMOS-integrated quantum diamond biosensing platform that combines nitrogen–vacancy (NV) centers in diamond with a custom 40 nm CMOS Single-Photon Avalanche Diode (SPAD) array. The system integrates on-chip active quenching and digital readout with external FPGA-based photon counting, compact microwave delivery, and practical optical excitation and collection schemes to support widefield optically detected magnetic resonance (ODMR).

System-level design considerations spanning fluorescence collection efficiency, detector count-rate capability, and microwave homogeneity are analyzed with biological compatibility and scalability in mind. Using superparamagnetic iron oxide nanoparticle (SPION)–labeled HEK293T cells as a representative use case, simple dipole-field estimates indicate that sub-\textmu T sensitivity is required to resolve ODMR shifts within typical ensemble linewidths. Based on the proposed architecture and efficiency analysis, a magnetic field sensitivity of approximately 90 nT/$\sqrt{\mathrm{Hz}}$ per pixel is estimated.

These results outline a practical path from optics-heavy quantum diamond microscopes toward compact, CMOS-integrated NV-based biosensors for quantitative magnetic imaging in complex biological environments.
\end{abstract}


\titlepgskip=-15pt

\maketitle

\section{Introduction}
\label{sec:introduction}
\PARstart{S}{ensing} platforms are indispensable to biological and medical sciences, facilitating detailed investigations into complex biological structures and processes. Traditional bio-imaging methods, including optical, electron, and fluorescence microscopy, often face limitations such as autofluorescence and scattering, reducing image quality in complex biological samples \cite{glennSinglecellMagneticImaging2015}. Unlike conventional optical methods, magnetic field sensing is inherently immune to scattering and autofluorescence, since magnetic signals are not affected by optical absorption or emission processes, enabling robust imaging in complex biological environments. Quantum diamond microscopy (QDM) addresses these challenges by exploiting the quantum properties of nitrogen-vacancy (NV) centers in diamond, which encode local magnetic fields in optically detectable resonance shifts, providing a quantum-enabled alternative for biological imaging \cite{hallHighSpatialTemporal2012, glennSinglecellMagneticImaging2015}.

At the core of this capability, the NV center's spin states are sensitive to external perturbations such as magnetic fields, electric fields, temperature variations, and mechanical strain \cite{mazeNanoscaleMagneticSensing2008, doldeElectricfieldSensingUsing2011, acostaTemperatureDependenceNitrogenVacancy2010}. Such sensitivity, combined with exceptional photostability and biocompatibility, makes NV centers versatile quantum sensors suitable for diverse applications, particularly in biomedical imaging and diagnostics \cite{mochalinPropertiesApplicationsNanodiamonds2012, glennSinglecellMagneticImaging2015}.

\Figure(topskip=0pt, botskip=0pt, midskip=0pt)[width=\textwidth]{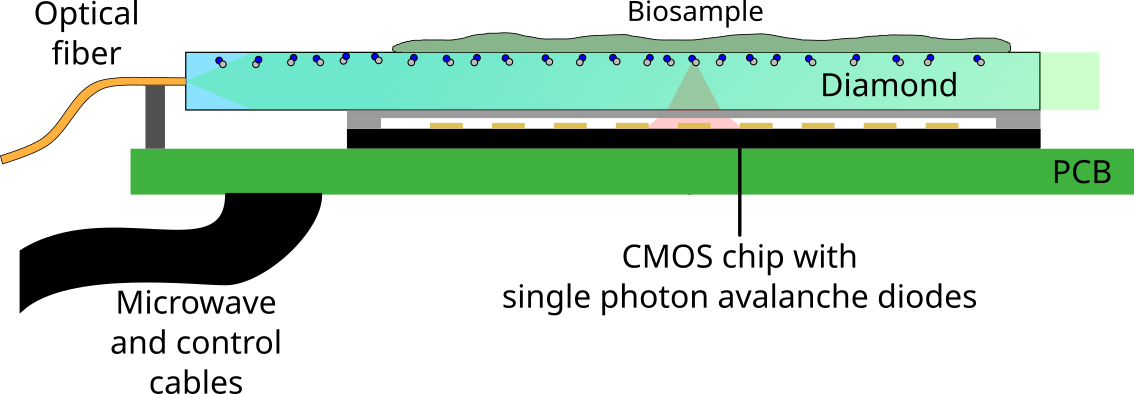}
{Concept design of an integrated diamond-on-chip NV-based quantum biosensor.\label{fig:integrated_SPAD_NV_Biosample}}

These properties have motivated sustained research into NV centers over several decades, beginning in the mid-1960s \cite{preezElectronParamagneticResonance1965} and accelerating following the first detection of single negatively charged NV centers in 1997 \cite{gruberScanningConfocalOptical1997}. Since then, NV centers have become prominent in quantum information processing and quantum metrology, with demonstrated nanoscale sensitivity suitable for advanced medical diagnostics and enhanced magnetic resonance imaging (MRI) \cite{maletinskyRobustScanningDiamond2012, maminNanoscaleNuclearMagnetic2013, balasubramanianNanoscaleImagingMagnetometry2008, rugarProtonMagneticResonance2015, borettiSingleBiomoleculeImaging2015, zalieckasQuantumSensingMicroRNAs2024}. Their biocompatibility and stability at room temperature further highlight their suitability for bio-imaging \cite{mochalinPropertiesApplicationsNanodiamonds2012, balasubramanianUltralongSpinCoherence2009}.

Historically, NV-based sensing setups have been bulky and costly, due to their reliance on free-space optics, external microwave hardware, and discrete detection electronics. Recent developments, however, emphasize miniaturization and integration with complementary metal-oxide-semiconductor (CMOS) technologies, drastically reducing system size, cost, and complexity \cite{webbNanoteslaSensitivityMagnetic2019, kuwahataMagnetometerNitrogenvacancyCenter2020, sturnerIntegratedPortableMagnetometer2021, wangPortableDiamondNV2022}. Innovations include portable NV-based sensors with compact USB-driven designs \cite{deguchiCompactPortableQuantum2023} and integrated LED-based excitation methods \cite{pogorzelskiCompactFullyIntegrated2024}. CMOS-integrated NV sensors have established the feasibility of compact, integrated quantum sensing solutions \cite{kimCMOSintegratedQuantumSensor2019, ibrahimHighScalabilityCMOSQuantum2021}.

In this research, we aim to advance integration by addressing the need for scalable, on-chip detection of weak NV fluorescence with high temporal resolution by employing Single-Photon Avalanche Diodes (SPADs). In addition to seamless compatibility with advanced CMOS processes, SPADs provide picosecond timing resolution, high detection efficiency across the NV emission band (600–800 nm), and scalability into dense arrays. These features make SPADs well suited for quantum biosensing, where detecting weak, rapidly varying fluorescence signals from NV ensembles is essential \cite{levinePrinciplesTechniquesQuantum2019}. Our proposed architecture combines a CMOS-based SPAD array with an NV-containing diamond substrate, using on-chip photonic waveguides for efficient optical excitation. Biosamples can be directly attached to the diamond surface, minimizing sensor-to-sample distances and optimizing sensitivity (\autoref{fig:integrated_SPAD_NV_Biosample}).

This integrated NV–SPAD platform is expected to significantly enhance bio-imaging speed and spatial resolution, offering a powerful tool for biomedical diagnostics and fundamental research. By combining advanced CMOS detector capabilities with the unique spin-dependent fluorescence of NV centers, the platform directly links toward integrated photonics with practical biosensing applications.

Integrated NV sensors using this setup are scalable, since the sensor is a stack of 2D components that can be arbitrarily extended with more pixels, the only limit being the bandwidth of the chip-to-computer connection. This kind of system requires CMOS fabrication and micron-scale assembly, both of which are feasible to setup for mass production. In biosensing, resolving details at the cellular level requires a spatial resolution in the order of microns. Such details are crucial for cancer detection. Besides that, succesful NV magnetometry has been demonstrated for cardiography \cite{yuNoninvasiveMagnetocardiographyLiving2024} and neuron imaging \cite{hallHighSpatialTemporal2012}. For cancer detection, Magnetic Resonance Imaging (MRI) is a powerful diagnostic tool, but suffers from limited spatial resolution (e.g. 60 micron at 9.4 T \cite{chenImmunomagneticMicroscopyTumor2022}) and is bulky and costly. The proposed integrated NV wide-field magnetometry has potential for much lower spatial resolution, ultimately limited by the pixel pitch.

This paper is organized as follows. Section~II introduces the physical principles of nitrogen-vacancy (NV) centers and optically detected magnetic resonance (ODMR) that underpin quantum biosensing. Section~III presents the design and operation of the CMOS-integrated SPAD array. Section~IV describes the digital readout and control architecture, while Section~V details the microwave delivery strategy required for ODMR. Section~VI discusses excitation and fluorescence collection schemes and evaluates their efficiency. Section~VII provides a sensitivity analysis and system-level performance estimation. Section~VIII outlines the proposed device architecture and key characteristics. Section~IX demonstrates the biosensing use case using SPION-labeled HEK293T cells. Finally, Section~X concludes the paper and outlines directions for future work.

\section{NV Centers \& ODMR Principles}
\label{sec:nv_odmr}

Nitrogen-vacancy (NV) centers in diamond are point defects consisting of a nitrogen atom adjacent to a vacancy site, substituting a carbon atom in the diamond lattice \cite{gruberScanningConfocalOptical1997, dohertyTheoryGroundstateSpin2012}. These centers exhibit remarkable quantum sensing properties due to their electronic spin states, primarily characterized by transitions between the spin-triplet ground state ($^3A_2$) and the excited state ($^3E$), with intersystem crossings via intermediate singlet states ($^1A_1$, $^1E$) \cite{dohertyTheoryGroundstateSpin2012}. 

In practical sensing implementations, however, not all NV charge states contribute equally to spin-dependent optical readout. While both neutral (NV$^0$) and negatively charged (NV$^-$) states exist, only NV$^-$ exhibits spin-dependent fluorescence and long spin coherence, whereas NV$^0$ lacks a stable spin ground state, making NV$^-$ the focus of quantum sensing \cite{gruberScanningConfocalOptical1997, dohertyTheoryGroundstateSpin2012}.

NV$^-$ centers exhibit a zero-field splitting (ZFS) of approximately 2.87 GHz between their spin sublevels ($m_s = 0$ and $m_s = \pm1$) \cite{dohertyTheoryGroundstateSpin2012}. Upon optical excitation with green laser light (typically 532 nm), NV centers fluoresce predominantly in the red and near-infrared spectral range (600–800 nm). The fluorescence intensity depends significantly on the spin state, as non-radiative intersystem crossings preferentially occur from the excited $m_s = \pm1$ states to singlet states, resulting in reduced photoluminescence compared to the $m_s = 0$ state \cite{gruberScanningConfocalOptical1997, jelezkoSingleDefectCentres2006}. This spin-dependent fluorescence forms the fundamental mechanism for optically detected magnetic resonance (ODMR) \cite{gruberScanningConfocalOptical1997, jelezkoObservationCoherentOscillations2004}.

\begin{figure*}[!t]
    \centering
    \includegraphics[width=0.8\textwidth]{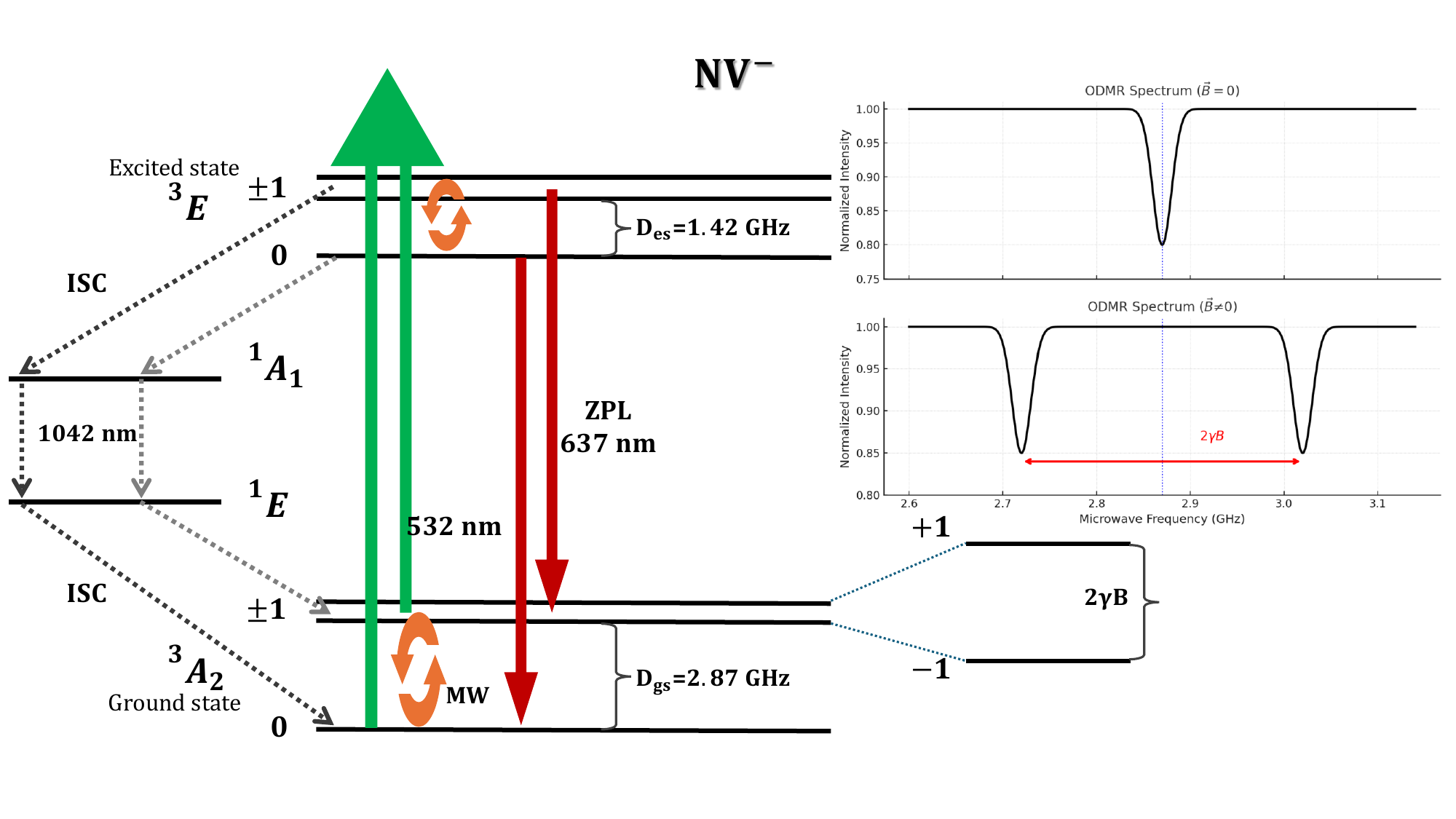}
    \caption{Energy-level structure and ODMR principle of the negatively charged nitrogen-vacancy (NV$^-$) center. The ground and excited spin-triplet states exhibit a zero-field splitting of approximately 2.87~GHz between the $m_s = 0$ and $m_s = \pm1$ sublevels. Optical excitation (typically at 532~nm) induces spin-dependent fluorescence in the 600–800~nm range via intersystem crossings through intermediate singlet states. The right panel illustrates representative continuous-wave ODMR spectra: at zero magnetic field ($B = 0$), the $m_s = \pm1$ transitions are degenerate, while an applied magnetic field ($B \neq 0$) lifts this degeneracy through Zeeman splitting, resulting in two distinct resonance frequencies separated proportionally to the magnetic field component along the NV axis.}
    \label{fig:NV_energy_levels}
\end{figure*}

ODMR leverages microwave (MW) fields resonant with the NV center’s electron spin transitions to induce changes in fluorescence intensity, enabling sensitive detection of magnetic fields. Application of an external magnetic field causes Zeeman splitting, lifting the degeneracy of the $m_s = \pm1$ states. This produces two resonance frequencies separated by $\Delta f \approx 2 \gamma B_\parallel$, where $\gamma \approx 28$ GHz/T is the electron gyromagnetic ratio and $B_\parallel$ is the field component along the NV axis \cite{mazeNanoscaleMagneticSensing2008, balasubramanianNanoscaleImagingMagnetometry2008}. By measuring this splitting, ODMR enables quantitative magnetic field sensing with high sensitivity and spatial resolution.

Two primary ODMR approaches exist: continuous-wave (CW) and pulsed ODMR. CW-ODMR involves continuous optical and microwave excitation, providing straightforward implementation and stable measurements suitable for static or slowly varying fields \cite{jelezkoSingleDefectCentres2006}. Pulsed ODMR, employing sequences of timed optical and MW pulses, offers improved sensitivity and resolution for dynamic fields and advanced sensing protocols \cite{taylorHighsensitivityDiamondMagnetometer2008, barrySensitivityOptimizationNVDiamond2019}. For many biomedical applications, particularly those involving ensemble NV centers, CW-ODMR remains the preferred method due to its simplicity, rapid imaging capabilities, and robust performance in complex biological environments \cite{glennSinglecellMagneticImaging2015, hallHighSpatialTemporal2012}. In integrated platforms, the spin-dependent fluorescence can be collected with single-photon detectors such as CMOS-integrated SPAD arrays, enabling direct electrical readout of ODMR signals for widefield bioimaging \cite{levinePrinciplesTechniquesQuantum2019}.

\section{CMOS SPAD Array Design}
\label{sec: CMOS SPAD Array Design}

This section focuses on the detector requirements for NV-based quantum biosensing and explains why CMOS-integrated Single-Photon Avalanche Diode (SPAD) arrays are used in this work. It first reviews existing SPAD technologies and CMOS implementations, and then introduces the design approach for a compact SPAD array intended for direct detection of NV fluorescence and integration with a diamond sensing substrate.

\subsection{Background and State-of-the-Art}
\label{subsec: Background - SPADs}

Single-Photon Avalanche Diodes (SPADs) are a specialized class of Avalanche Photodiodes (APDs) that operate in Geiger mode, i.e., reverse-biased above their breakdown voltage. In this regime, the absorption of a single photon can trigger a self-sustaining avalanche of charge carriers, resulting in a detectable current pulse. This unique capability provides shot-noise-limited sensitivity to single photons and sub-nanosecond timing resolution, making SPADs highly attractive for a wide range of applications, including wide-field fluorescence lifetime imaging microscopy (FLIM) with megapixel SPAD cameras \cite{zickusFluorescenceLifetimeImaging2020}, time-of-flight (ToF) and depth imaging using time-gated SPAD sensors \cite{morimotoMegapixelTimegatedSPAD2020}, LiDAR and 3D imaging with CMOS SPAD arrays \cite{henderson$192times128$TimeCorrelated2019}, and quantum communication protocols where SPADs are widely employed as single-photon detectors \cite{hadfieldSinglephotonDetectorsOptical2009, zhangAdvancesInGaAsInP2015}. A broad overview of biophotonics applications is given in \cite{bruschiniSinglephotonAvalancheDiode2019}.

Over the past two decades, there has been increasing interest in the integration of SPADs with complementary metal-oxide-semiconductor (CMOS) technology. CMOS-SPAD integration offers several advantages, including higher array density and fill factor, scalability to large detector arrays, cost-effective mass fabrication, and reduced power consumption enabled by smaller transistor dimensions \cite{zappaPrinciplesFeaturesSinglephoton2007, leeHighPerformanceBackIlluminatedThreeDimensional2018, henderson$192times128$TimeCorrelated2019}. These features have made CMOS-integrated SPADs a cornerstone of modern single-photon detection platforms.

Significant progress has been achieved in the design of SPAD arrays implemented in advanced CMOS nodes. In 2017, Pellegrini \emph{et al.} \cite{pellegriniIndustrialisedSPAD402017} demonstrated one of the first mature SPAD arrays fabricated in 40~nm CMOS technology. Their prototype featured 64 SPADs organized into four macropixels, each containing 16 SPADs with dedicated readout electronics. In parallel, Al Abbas \emph{et al.} \cite{alAbbasGlobalSharedWellSPAD2017} developed a $96 \times 40$ SPAD array on a compact $1 \times 1$ mm$^2$ chip, and subsequently extended their work to a $192 \times 128$ array fabricated using STMicroelectronics' 40 nm CMOS process \cite{henderson$192times128$TimeCorrelated2019}. This achievement represented a major step forward in pixel density and scalability. More recently, Morimoto \emph{et al.} \cite{morimoto32Megapixel3DStacked2021} reported a 3.2\,megapixel SPAD array, realized using 3D-stacking techniques and a different CMOS technology node. To date, this remains the highest-resolution SPAD image sensor publicly reported.

Although most work has focused on 40 nm technology, exploratory studies have also investigated SPAD fabrication and optimization in 28 nm CMOS \cite{deAlbuquerqueIntegrationSPAD2018, issartelArchitectureOptimizationSPAD2022, gaoCorrelationsDCRPDP2023}. These efforts highlight the rapid evolution of SPAD technology toward smaller nodes, higher integration density, and enhanced fill factor, all of which are critical for next-generation applications such as integrated quantum sensing.

In the context of this work, the goal is to extend this progress by integrating a CMOS-fabricated $16 \times 16$ SPAD array, equipped with on-chip quenching and readout circuits, directly with a diamond substrate containing a doped layer of nitrogen-vacancy (NV) centers. This co-integration leverages the maturity of CMOS-SPAD technology while enabling direct detection of NV fluorescence, an important step toward scalable diamond-based quantum biosensors.

\subsection{Operating Principles}
\label{subsection: Operating Principles}

Single-Photon Avalanche Diodes (SPADs) are based on p–n junctions operated in Geiger mode, i.e., reverse-biased above their breakdown voltage. In this regime, the absorption of a single photon generates an electron–hole pair, which is accelerated by the high electric field, initiating a chain reaction of carrier multiplication. This self-sustaining avalanche produces a measurable current pulse that signals the detection of a photon \cite{covaAvalanchePhotodiodesQuenching1996}.

\begin{figure}[h]
    \centering
    \includegraphics[width=\linewidth]{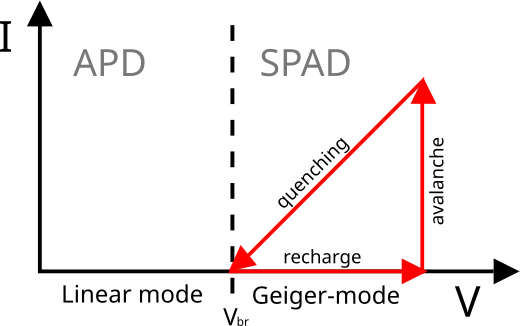}
    \caption{I-V characteristic of photodiodes. The conventional Avalanche Photo Diode (APD) gives an analog signal that is amplified by incoming light. The Single Photon Avalanche Diode (SPAD) is operated in reverse bias beyond breakdown voltage $V_{br}$, pushing it in the Geiger regime. An incoming photon triggers an avalanche, which becomes self-sustaining. Then, either passive or active quenching can be used to pull the voltage down and stop the avalanche. Lastly, the SPAD is allowed to recharge, making it ready for the next photon detection.}
    \label{fig:typical_IV}
\end{figure}

After each avalanche event, the device must be reset before detecting another photon. This is accomplished through quenching, which reduces the bias voltage below breakdown to stop the avalanche. Two main quenching strategies exist. In \emph{passive quenching}, a ballast resistor in series with the SPAD limits the avalanche current, lowering the junction voltage below breakdown. The voltage then recovers gradually, restoring the SPAD to its operating point. While simple and cost-effective, passive quenching is slow and susceptible to higher afterpulsing \cite{covaAvalanchePhotodiodesQuenching1996}. In contrast, \emph{active quenching} employs dedicated circuitry to sense the avalanche current and quickly pull the bias below breakdown. This enables rapid quenching, followed by a controlled recharge, and allows precise control over the dead time between detection events—an important parameter for high-speed and low-afterpulsing operation \cite{dautetPhotonCountingTechniques1993}. \autoref{fig:typical_IV} illustrates the operating principle of SPADs compared to standard avalanche photodiodes (APDs).

\subsection{Design Considerations for NV Sensing}
\label{subsection: SPAD Parameters}

The suitability of SPADs for NV-based biosensing is determined by several performance parameters:

\begin{itemize}
   \item \textbf{Photon Detection Efficiency (PDE)} — the probability that an absorbed photon initiates an avalanche. For NV fluorescence detection, a peak PDE in the 600–800~nm band is required \cite{zappaPrinciplesFeaturesSinglephoton2007, charbonSinglephotonImagingComplementary2014}.

   \item \textbf{Timing Resolution} — SPADs can achieve timing jitter on the order of hundreds of picoseconds (e.g., $\sim$150~ps FWHM), which is sufficient for NV-based sensing where fluorescence lifetimes are on the order of several nanoseconds and time-correlated single-photon counting (TCSPC) can be used to reject background and improve signal contrast \cite{stipcevicCharacterizationNovelAvalanche2010}.
     
    \item \textbf{Dark Count Rate (DCR)} — spurious counts arising from thermally generated carriers or defects. For NV ensemble fluorescence detection, a DCR on the order of a few kHz or lower is desirable \cite{zappaPrinciplesFeaturesSinglephoton2007}.
    
    \item \textbf{Afterpulsing} — avalanches triggered by trapped carriers released after an initial event. For NV-based sensing, afterpulsing must be kept sufficiently low to avoid artificial count correlations on microsecond timescales, which can distort photon statistics and reduce ODMR contrast. This can be suppressed with optimized doping profiles and active quenching circuits incorporating programmable dead time \cite{stipcevicCharacterizationNovelAvalanche2010}.
\end{itemize}

The SPADs in this work are implemented in TSMC’s 40~nm CMOS technology. This node provides high integration density, reduced power consumption, and compatibility with complex on-chip electronics. Each device must sustain photon detection rates above 3~Mcounts/s (corresponding to a cycle time below 333~ns) to prevent signal bottlenecks when measuring ensemble NV emission. Crosstalk is a concern at small pixel pitches (here $\sim$20~$\mu$m), but can be mitigated by careful layout and shielding. To suppress afterpulsing and ensure fast recovery, the quenching time is targeted at below 20~ns \cite{ghioniCompactActiveQuenching1996}. Finally, the 40~nm process constrains transistor operation to a nominal supply voltage of 1.1~V, which defines the design envelope for on-chip circuitry.

\subsection{Chip Architecture with Integrated Metal Grating}
\label{subsection: Chip specifications}

The SPAD array designed for this work comprises a $16 \times 16$ pixel grid, equipped with on-chip quenching, recharge, and readout circuitry. The array size was selected as the maximum number of SPAD pixels that could be integrated, together with their associated circuitry described below, within the available chip area of slightly less than 1~mm$^2$. The design is compatible for grating filters to be integrated on top of the SPADs, such as how Ibrahim et al did it \cite{inproceedings}. Such a filter suppresses the intense 532~nm green excitation light while transmitting the desired NV fluorescence in the 600–800~nm band. The grating pitch and duty cycle were optimized to produce destructive interference at the pump wavelength and constructive transmission across the fluorescence band, consistent with grating interference filtering principles. This integration provides an on-chip optical filter that enhances the signal-to-background ratio without the need for bulky external optics.

\begin{figure}[h!]
    \centering
    \includegraphics[width=\linewidth]{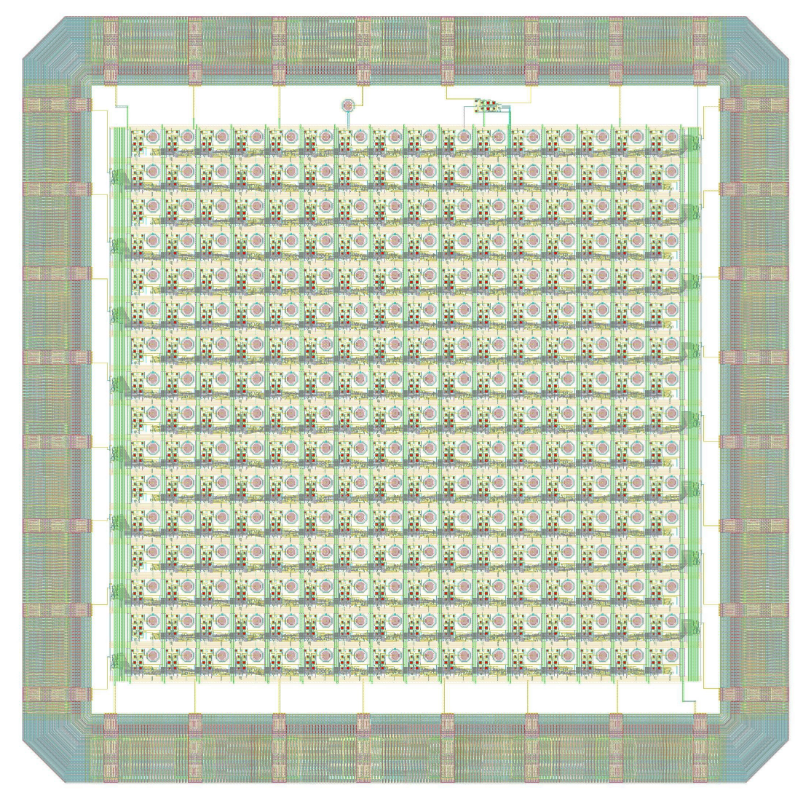}
    \caption{Layout of the SPAD chip with a $16 \times 16$ array, with on-chip active quenching and readout electronics.}
    \label{fig:full_chip}
\end{figure}

The overall chip layout was implemented using Cadence Virtuoso. Each die occupies an area of $1085 \times 1085$~$\mu$m$^2$ in design, reduced by a shrink factor of 0.9 during fabrication, yielding final dimensions of $976.5 \times 976.5$~$\mu$m$^2$. \autoref{fig:full_chip} shows the top-level layout of the chip.

Beyond the physical layout, the functionality of the chip is defined by its hierarchical organization, which ensures efficient event collection and communication from individual pixels to the periphery.

\subsection{Chip Design and On-Chip Circuitry}
\label{subsection: Chip design and circuitry}

The architecture is organized hierarchically, from SPAD circuits, to pixels, up to the top-level bus structure, enabling efficient serialization and transmission of photon events. The overall chip structure, shown in \autoref{fig:full_chip}, comprises 16 rows of SPAD units, each connected to a global bus. Every row incorporates a 16-bit Parallel-In-Serial-Out (PISO) circuit that serializes photon detection events for transmission off-chip. Because the array far exceeds the 32 available I/O pins, this serialization step is essential to avoid data loss. In addition, two 5-bit Serial-In-Parallel-Out (SIPO) modules configure the timing parameters of the hold-off circuit, specifically controlling the width and delay of the recharge pulse. A common bus distributes the applied bias voltage $V_{bias}$, the global clock, the configuration bits, and the reset signal across all rows, ensuring synchronized operation throughout the array.

\begin{figure*}
    \centering
    \includegraphics[width=\textwidth]{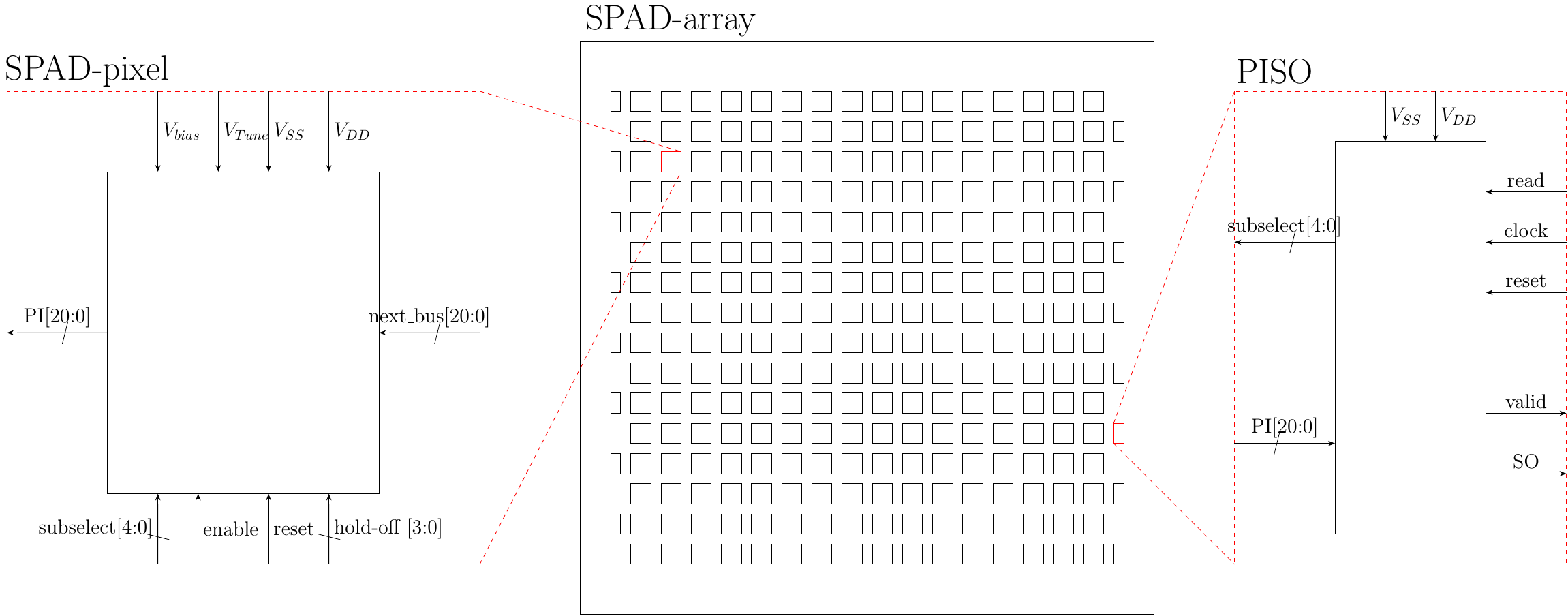}
    \caption{Overview of the SPAD array. The array consists of 16x16 SPAD pixels, which are organized into rows. The pixel schematic is highlighted on the left. Each row has a Parallel In Serial Out (PISO) module (highlighted on the right), which makes the pixels accessible with their respective serial I/O.}
    \label{fig:SPAD-array-virtuoso}
\end{figure*}

\begin{figure}
    \centering
    \includegraphics[width=\linewidth]{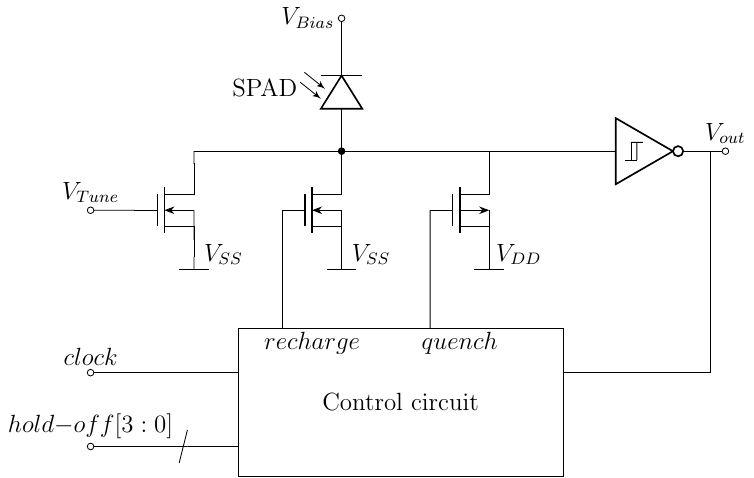}
    \caption{Schematic of the SPAD circuit. The active quenching, hold-off and recharge is managed by the control circuit, which pulls the anode to $V_{DD}$ and $V_{SS}$ accordingly.}
    \label{fig:SPAD-circuit_schematic}
\end{figure}

Each SPAD is equipped with a readout circuit, as shown in \autoref{fig:SPAD-circuit_schematic}. This circuit handles the quenching, holdoff and recharge of the SPADs. A Schmidt trigger is used to transform the SPAD pulses in reliable digital pulses. The series resistance can be tuned with the voltage $V_{tune}$. This way, the system can be corrected for different avalanche currents. The required resistance is typically in the order of 10 - 100 k$\Omega$. The applied bias voltage $V_{bias}$ ensures Geiger-mode operation of the SPADs. When an avalanche is triggered, the control circuit quenches the SPAD. Next, it waits for a programmable hold-off time, before resetting the SPAD by recharging it. The hold-off limits afterpulsing, which is beneficial for overall sensor sensitivity.

\begin{figure}
    \centering
    \includegraphics[width=\linewidth]{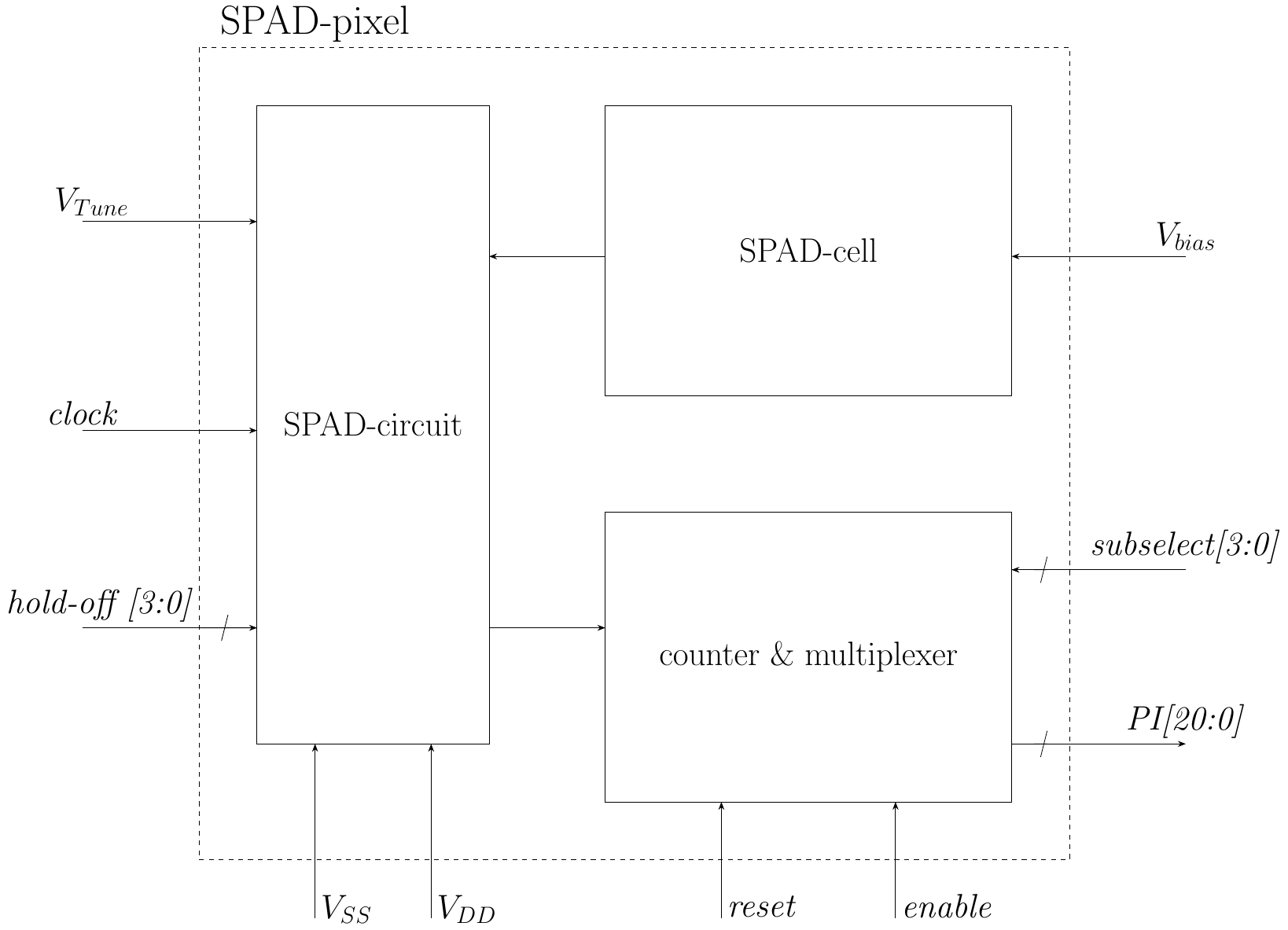}
    \caption{Layout of the SPAD pixel. "SPAD-cell" refers to the actual SPAD itself, and "SPAD-circuit" represents the surrounding circuitry. "SPAD-cell" and "SPAD-circuit" together represent the circuit from \autoref{fig:SPAD-circuit_schematic}. Besides that, a counter and multiplexer are included that count the pulses and communicate them with the PISO module.}
    \label{fig:SPAD-pixel_tikz}
\end{figure}

\begin{figure}
    \centering
    \includegraphics[width=\linewidth]{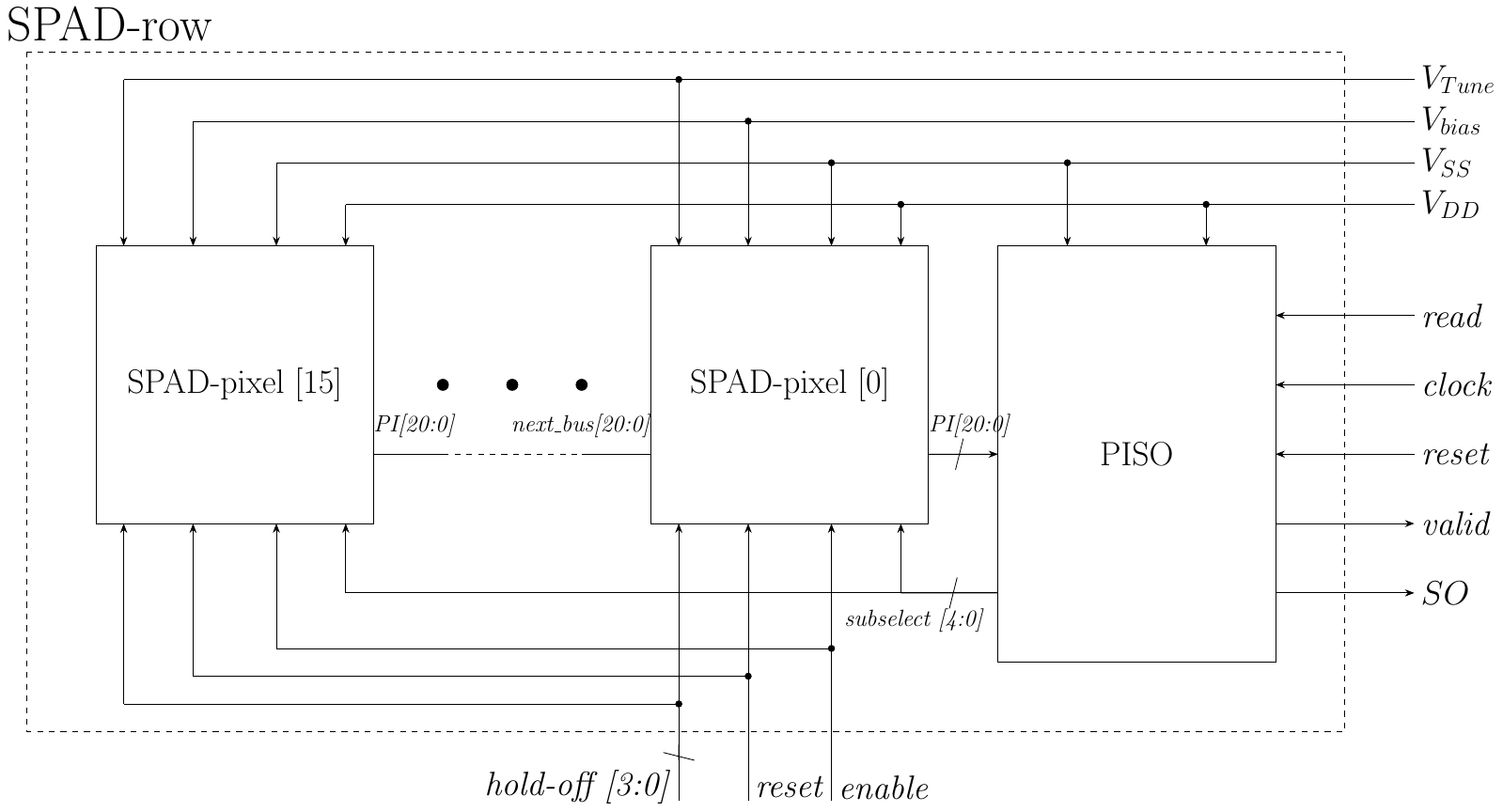}
    \caption{This schematic shows how one row of SPAD pixels are connected to eachother and the PISO module. Ultimately, the data of the pixels is serialized, and an FPGA reads the serial data of each row by interpreting the data sent through Serial Out (SO).}
    \label{fig:SPAD_row}
\end{figure}

At the pixel level, the SPAD and its accompanying circuit are connected with a counter and multiplexer module, as depicted in \autoref{fig:SPAD-pixel_tikz}. The circuit from \autoref{fig:SPAD-circuit_schematic} is represented by blocks "SPAD-cell" and "SPAD-circuit" in \autoref{fig:SPAD-pixel_tikz}. The counter \& multiplexer module counts the digital pulses, from $V_{out}$, where each pulse represents a photon event. The multiplexer serves a double purpose; The pixel either sends it value if it's the selected pixel, or it routes the command to the next pixel. This approach allows the pixels to be chained together, reducing the amount of centrally connected wires needed to connect pixels, thus improving scalability. The configuration of these pixel rows is depicted in \autoref{fig:SPAD_row}.

The fabricated die, implemented in the TSMC 40~nm CMOS process, integrates the SPAD array together with the active quenching and hold-off circuitry. Building on the physical layout described earlier, the focus here is on the pixel-level electronics that enable high-speed operation. Each SPAD unit incorporates an active quenching circuit that detects avalanche onset, pulls the bias below breakdown, and restores the operating voltage. This eliminates the need for bulky passive resistors and enables high count rates. While our design targets quench times of 20~ns and recharge within 1~ns (corresponding to $\sim$40~Mcounts/s), recent state-of-the-art implementations have demonstrated active reset circuits with dead times as short as 50~ps, supporting count rates well into the gigacount regime \cite{dolatpoorlakehUltrafastActiveQuenching2021}. The implemented circuit, shown in \autoref{fig:SPAD-circuit_schematic}, consists of multiple transistors whose gates are controlled by a separate control circuit, which controls when the SPADs are quenched and recharged.

To further mitigate afterpulsing, each pixel includes a programmable hold-off circuit that digitally controls the dead time between successive avalanche events. This is part of the control circuit in \autoref{fig:SPAD-circuit_schematic}.
This digital approach provides precise programmability while maintaining compact layout and high-speed operation.

\begin{figure}
    \centering
    \includegraphics[width=\linewidth]{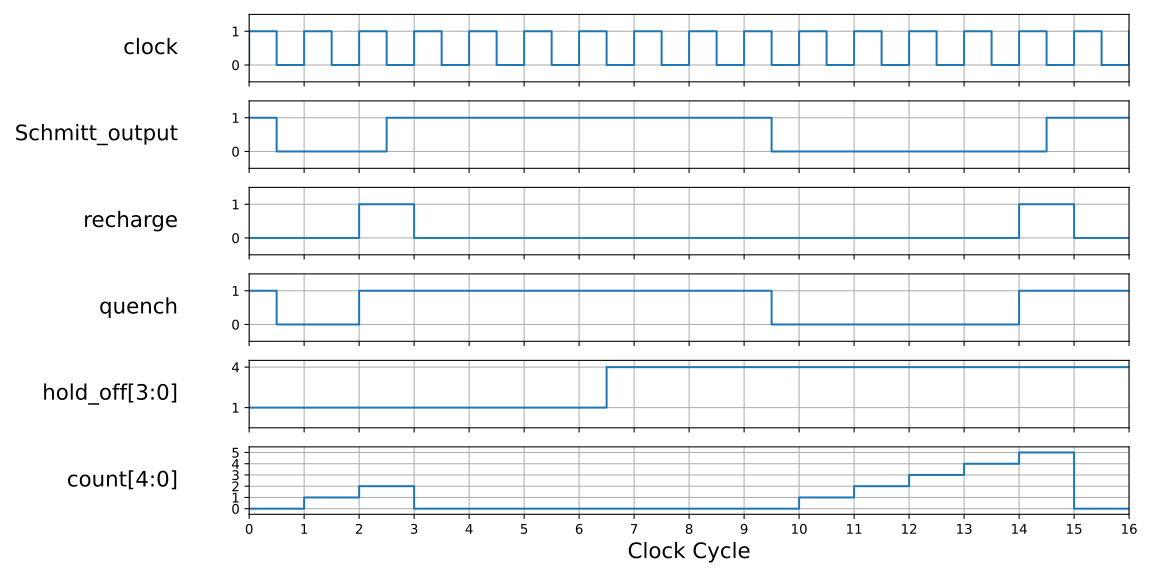}
    \caption{\textbf{Simulation of the SPAD-circuit}. The SPAD-circuit has been simulated with ModelSim with two different hold-off settings. The first setting is of value 1 and runs from cycle 0 to 6.5. The second setting is of value 4 and runs from cycle 6.5 to 16. From these two settings the difference in hold-off time can be observed and the behavior of quenching and recharge signals can be validated.}
    \label{fig:SPAD_chip_simulation}
\end{figure}

Finally, the post-layout validation flow included Design Rule Check (DRC), and a partial Layout Versus Schematic (LVS) verification that excludes the SPAD. SPADs from an earlier iteration of the chip validated at 20~ns cycle time per SPAD unit, including the AQC and hold-off circuitry, with negligible degradation. Moreover, serialization across a 16-pixel row achieved a 30~ns readout cycle. These values meet the speed and reliability requirements for NV fluorescence detection in ensemble-based biosensing applications, as shown in \autoref{fig:SPAD_chip_simulation}.

At the array level, several well-known challenges limit the scalability of SPAD-based detectors, including optical and electrical crosstalk, increased power consumption, and readout bandwidth constraints as pixel count increases. In the present $16 \times 16$ implementation, these effects remain limited due to the modest array size and per-pixel isolation, but they represent important considerations for future scaling to larger formats. Optical crosstalk is mitigated by pixel spacing and metal shielding, while electrical crosstalk is reduced through careful layout and local signal routing. Addressing these issues at larger scales will require additional architectural and process-level optimizations.

\section{Digital readout \& control}

The FPGA, Arduino, and lab computer form the digital backend required to process photon signals from the SPAD chip. The FPGA serves as a high-speed parallel counter and serializer, decoding the 16 serial data streams from the SPAD array. It aggregates photon counts and converts them into a slower cumulative format suitable for transfer to the Arduino and computer. Because of voltage mismatches between the chip logic (1.1~V) and the FPGA logic (lowest available setting is 1.5~V), voltage dividers were used to reduce the FPGA output voltage to approximately 1.085~V in order not to damage the chip. While functional, this approach introduced signal distortion and required lowering the FPGA operating frequency to 1~MHz to maintain signal integrity. The FPGA interprets voltages above 975~mV as logical highs, allowing for reading the chip's output without level shifting, though this narrow margin increases susceptibility to noise and fluctuations.

During operation, the FPGA continuously polls the SPAD chip by sending a trigger pulse that signals readiness to receive photon count data. The SPAD chip responds by transmitting all 256 pixel counters simultaneously across 16 serial channels. The serial pins through which this happens is depicted in \autoref{fig:SPAD_row}. The FPGA decodes these values, accumulates them in internal 32-bit registers, and then resets the on-chip counters to prepare for the next acquisition cycle. With 32-bit depth, the counters can store more than $4\times10^9$ events before overflow. Once the counts are aggregated, the FPGA transfers the data to the Arduino via a Serial Peripheral Interface (SPI). Acting as the SPI master, the FPGA sends data at a deliberately reduced clock rate—about three orders of magnitude lower than its 50~MHz system clock—so that the Arduino, operating sequentially at 16~MHz, has sufficient time to process incoming bytes. Each transfer consists of a 1~KiB buffer containing 256 counters (32~bits each), preceded by a 0xFFFFFFFF header that marks the start of a new frame. Importantly, data acquisition and transmission proceed concurrently: while one buffer is transmitted, the FPGA continues to accumulate new photon counts.

The Arduino acts as a bridge between the FPGA and the lab computer. It receives raw data frames via SPI and forwards them unaltered over USB. To simplify integration, our existing Arduino firmware was reused without modification. At the end of the chain, the lab computer receives the continuous data stream, detects the 0xFFFFFFFF frame header, and parses the subsequent 32-bit integers into a 2D array corresponding to the $16 \times 16$ SPAD array. This reconstructed photon count image can be visualized in real time and subsequently analyzed, including for ODMR measurements.

\begin{figure}[h]
    \centering
    \includegraphics[width=\linewidth]{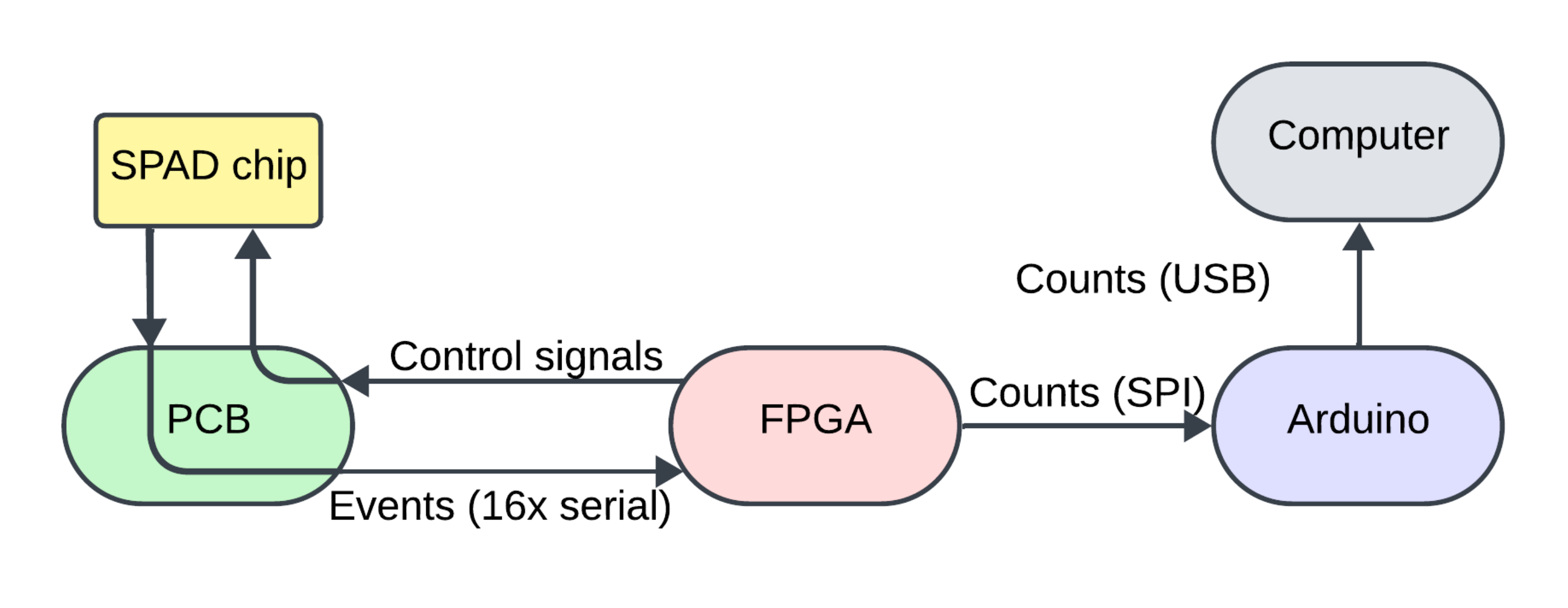}
    \caption{Communication chain between the SPAD chip, FPGA, Arduino, and lab computer.}
    \label{fig:schematic_SPAD-PC_comm_1}
\end{figure}

\begin{figure}[h]
    \centering
    \includegraphics[width=\linewidth]{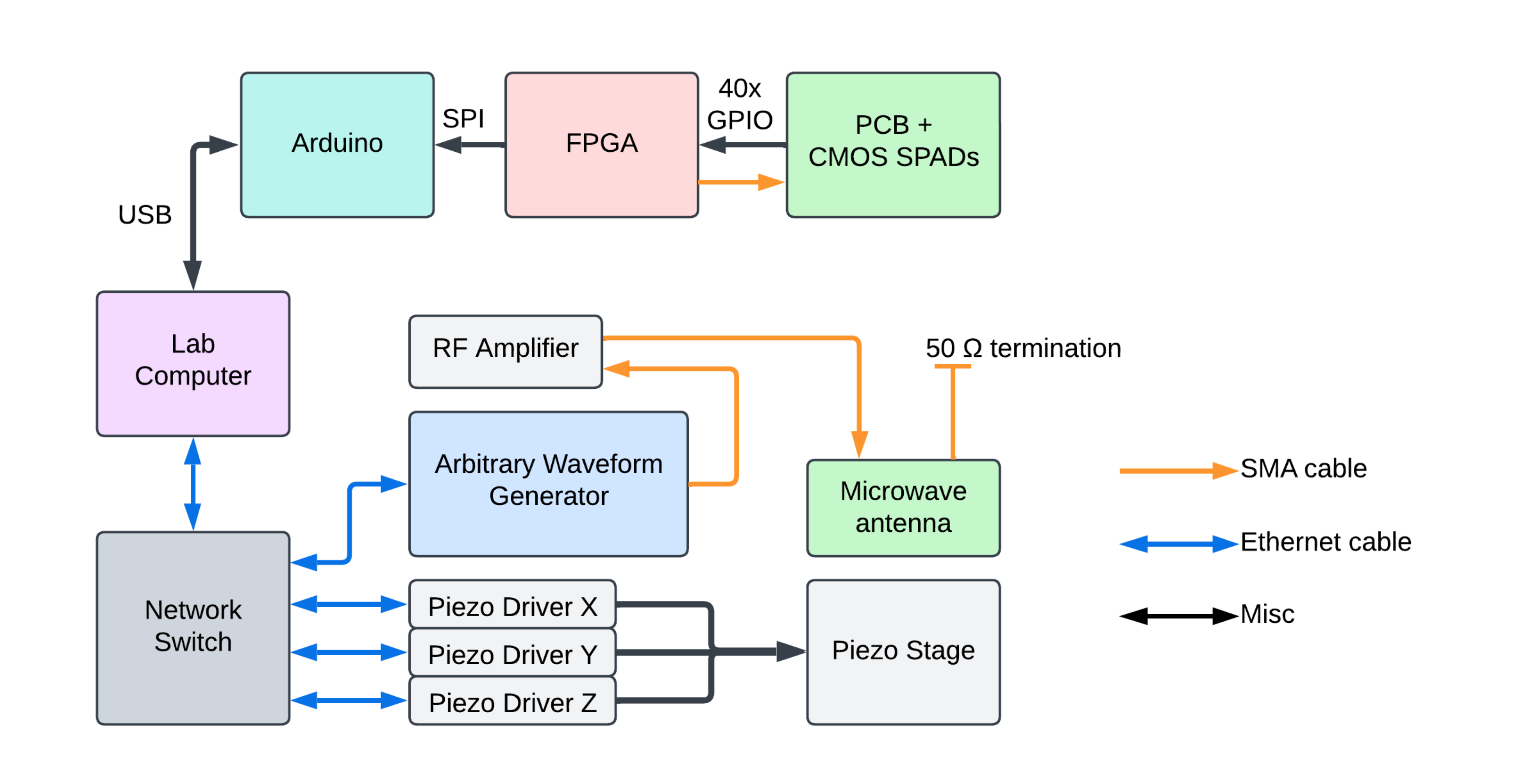}
    \caption{Experimental setup showing the SPAD chip, FPGA board, Arduino interface, and lab computer used for data acquisition.}
    \label{fig:Setup-and-communication_1}
\end{figure}


\section{Microwave Delivery}

With the digital backend established to capture spin-dependent fluorescence from the SPAD array, the final element required for ODMR operation is a microwave (MW) drive capable of uniformly addressing the NV ensemble. Efficient MW delivery is therefore a critical hardware component, with design requirements that include: (i) keeping conductive path lengths short at NV resonance frequencies ($\sim$2.87~GHz) to minimize phase accumulation and ohmic losses, (ii) generating homogeneous $B_1$ fields across the sensing region for widefield operation, and (iii) positioning the radiator close to the diamond to reduce power loss and field spreading.

At these frequencies (free-space wavelength $\sim$10~cm), traditional multi-turn coils become electrically long and suffer from interference and highly inhomogeneous $B_1$ distributions. Compact loop-based geometries, by contrast, remain viable when the conductor length of each element is kept short and placed close to the NV layer. Two practical families of antennas emerge for widefield sensing:

\textbf{Distributed loop arrays.} A grid of microfabricated single-turn loops positioned near the diamond surface can produce a relatively uniform $B_1$ field when driven coherently. Each loop contributes locally to the net field, while short current paths reduce phase errors and resistive losses. By adjusting the loop dimensions and spacing, the array can be tuned to balance field uniformity and drive efficiency.

\textbf{Parallel loop stacks.} Several small loops can be stacked or wired in parallel while keeping each loop’s perimeter much shorter than the MW wavelength. This configuration preserves the advantages of short path length and allows constructive superposition of $B_1$ fields. Unlike a conventional solenoidal coil, the stack avoids electrically long paths and suppresses standing-wave effects.

An alternative approach is to use a \textbf{grid of wires} patterned above or below the diamond \cite{ibrahimHighScalabilityCMOSQuantum2021}. With appropriate geometry, such a grid can serve simultaneously as a MW radiator and as an optical filter: it suppresses residual green pump light while transmitting red NV fluorescence. This dual-function architecture is especially attractive for compact widefield sensors, where optical filtering and MW delivery must coexist within a limited footprint.

For practical integration, the MW radiator is positioned within a few hundred micrometers of the NV layer, ideally separated only by the diamond itself and any required passivation or optical filter stack. This close spacing enhances drive efficiency, lowers the required MW power, and supports uniform Rabi oscillations across the sensing field. When diagonal or angled access is available (e.g., side-coupled or package-integrated structures), the radiator can be oriented to minimize optical shadowing while maintaining effective proximity to the NV layer.

\section{Excitation and Light Collection}
\label{sec:excitation_and_collection}

The optical subsystem completes the ODMR loop by providing two functions: (i) delivery of green excitation light to polarize the NV centers, and (ii) collection of spin-dependent red fluorescence for detection. The choice of excitation and collection scheme must balance efficiency, uniformity across the sensing field, and compatibility with both the SPAD array and the MW radiator.

\subsection{Excitation methods}

Three complementary excitation approaches are considered:

\subsubsection*{Objective-based excitation} Illumination through a high-numerical-aperture (NA) objective is straightforward in a laboratory setting. Top-side excitation through biological media can suffer from scattering and absorption, while bottom-side excitation alleviates these issues and allows the same objective to be used for fluorescence collection. Alternatively, a defocused wide beam can be used for excitation, removing the objective from the pump path while retaining it for fluorescence collection.

\subsubsection*{Total internal reflection (TIR)} Coupling a laser into the diamond at supercritical angles (via side coupling, diagonal bottom coupling, or fiber injection) generates an evanescent excitation sheet near the surface. This improves uniformity across the NV layer and allows compact packaging. Fiber-based delivery is alignment-stable and portable, making it well suited for deployable systems. Fiber-coupled NV magnetometers have been demonstrated \cite{liFibercoupledScanningMagnetometer2023}. TIR shallow-angle laser excitation has also been demonstrated before \cite{araiMillimetrescaleMagnetocardiographyLiving2022} \cite{glennSinglecellMagneticImaging2015}.

\subsubsection*{Waveguide-based excitation} Integrated photonic waveguides can route pump light on-chip and enhance light–matter interaction in single-NV or sparse-ensemble regimes. Such architectures, however, add fabrication complexity and demand tighter alignment tolerances with the external laser.

\subsection{Fluorescence collection}

NV fluorescence is emitted over a wide angular distribution, making efficient collection essential for maximizing signal-to-noise ratio. Below, multiple collection systems are outlined and compared.

\subsubsection*{Objective collection} A high-NA objective—often combined with bottom-side excitation—remains the standard in laboratory instruments. It provides strong collection efficiency and spatial resolution but increases size and alignment overhead. A bulky setup that needs alignment is less convenient to setup and use in the field. Besides that, objective lenses are generally expensive (up to a few thousand euros). Here, our aim is to present a compact integrated and economical alternative. Hence, below, alternatives are explored.

\subsubsection*{SPAD-under-diamond collection} Mounting the SPAD array beneath a thin diamond layer enables direct, widefield detection. Internal reflections at the diamond interfaces constrain the acceptance cone, limiting geometric collection to a few percent but simultaneously reducing pixel cross-talk. Microlenses fabricated on the diamond underside can increase coupling into SPAD pixels, though such relief structures may conflict with TIR-excitation schemes that require uninterrupted planar interfaces.

\subsubsection*{Waveguide collection} On-chip waveguides can preferentially route NV fluorescence in the 600–800~nm band while rejecting residual green pump light, improving background suppression. For fully integrated devices, waveguides ideally support both excitation and collection; otherwise, waveguide collection and TIR excitation may interfere unless carefully co-designed.

\subsection{Internal reflection stack concepts}

TIR-based excitation admits several optical stack designs. Assuming that the pixel spacing is sufficiently large to neglect inter-pixel optical overlap, we compare designs based solely on geometric (solid-angle) collection. Additional factors such as Fresnel reflection, scattering, and detector quantum efficiency are considered separately. Three representative designs are analyzed by numerically propagating rays, accounting for refraction and Fresnel transmission using classical optics. This is performed by integrating over surface elements of the diamond interface closest to the detector and evaluating how much emitted light exits through that surface and enters the detector. The chosen geometric values for these calculations are 10 \textmu m for each layer. The diamond, the pillar layer, and an air gap between detector and diamond each count as a layer for the designs where they apply.

\subsubsection*{The Simple Design - diamond slab over SPADs}
\begin{figure}
    \centering
    \includegraphics[width=\linewidth]{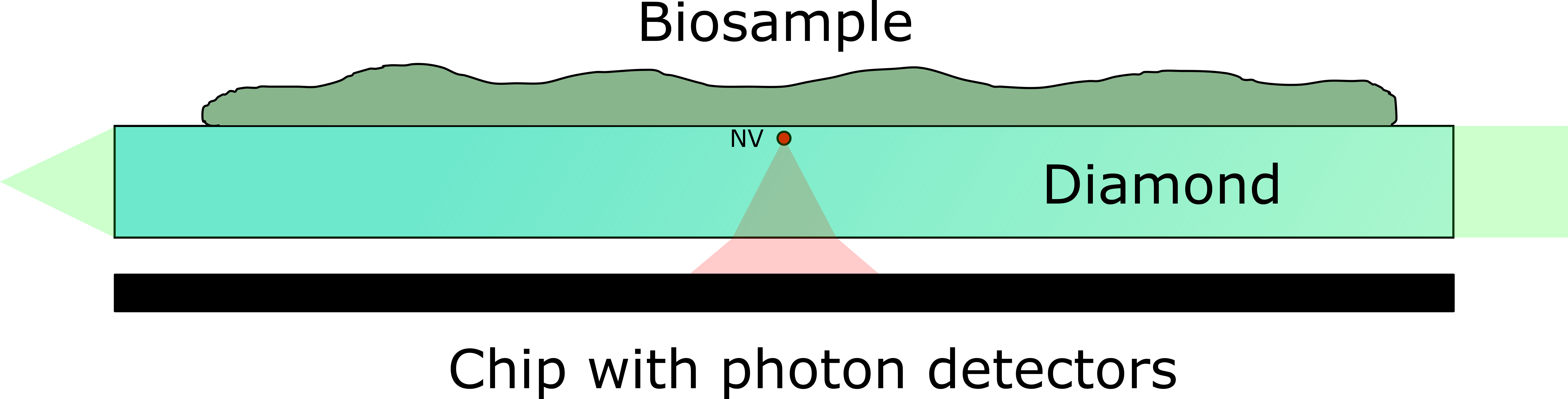}
    \caption{The simple design (diamond slab above SPADs) For a 10 \textmu m thick diamond, this design has a peak collection efficiency of 0.75\%. It collects from a wide region around it, collecting the equivalent of what 4.6 \textmu m $^2$ of diamond emits.}
    \label{fig:design_simple}
\end{figure}

A flat diamond slab positioned directly above the SPAD array collects only the emission cone subtended by each pixel at the NV layer (see \autoref{fig:design_simple}). Due to the absence of focusing, only a small fraction of the fluorescence reaches the SPADs. However, this configuration collects from an area substantially larger than the SPAD active area, ultimately leading to sufficient photon counts. Each SPAD collects fluorescence light most efficiently from NV centers located directly above it, with coupling decreasing toward zero for distant NV centers.

\begin{figure}
    \centering
    \includegraphics[width=\linewidth]{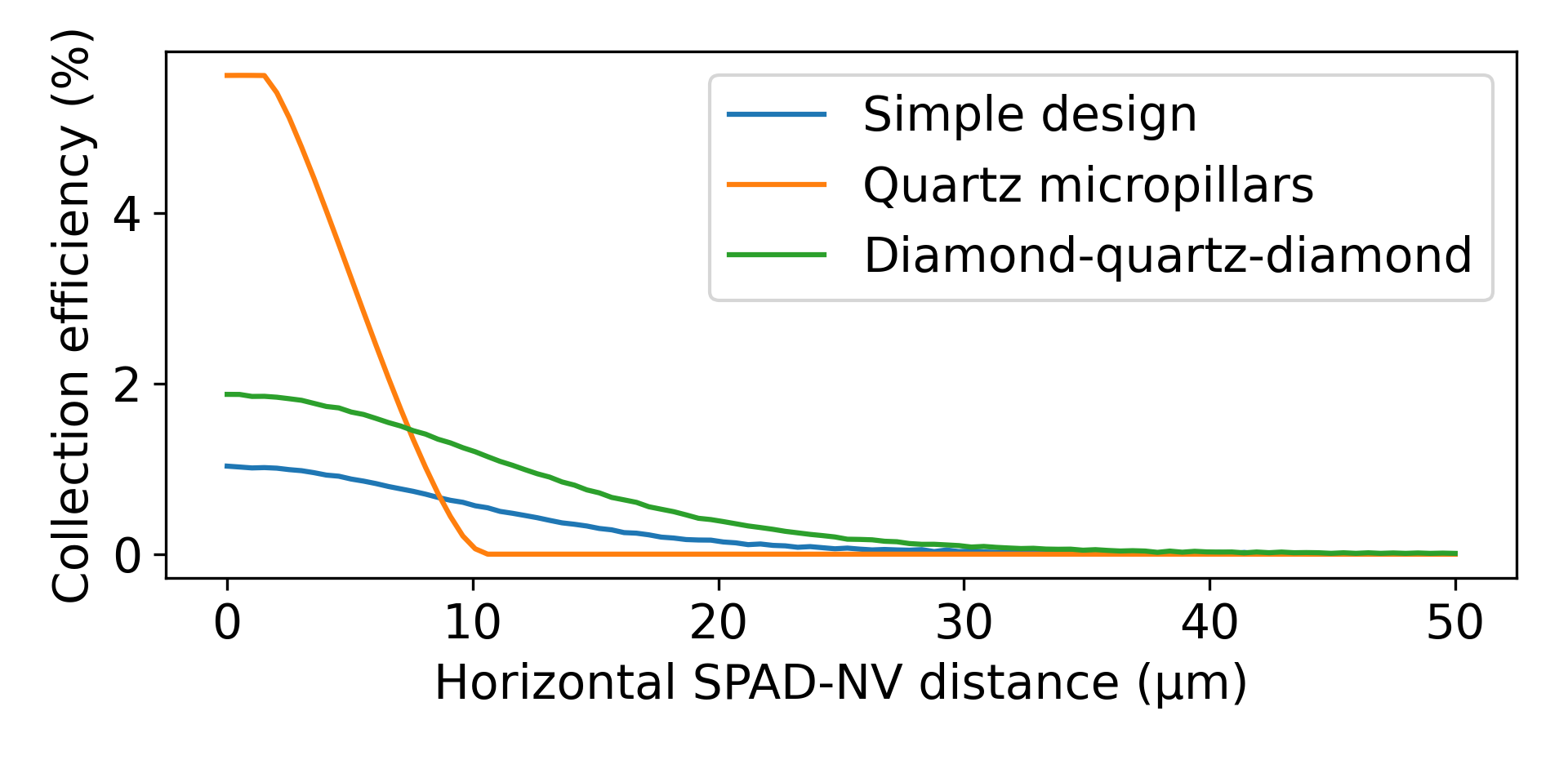}
    \caption{The fluorescence light collection distribution functions of the three discussed designs. These curves were numerically solved by integrating over the diamond surface nearest to the photodetectors and applying Snell and Fresnel equations to determine how much fluorescence light enters the sensor from any given NV position. The quartz micropillar has the best spatial selectivity. The diamond-quarts-diamond design has the same shape as the simple design, but generally collects more photons, doubling even that of the quartz micropillars. These calculated collection efficiencies take into account Fresnel losses.}
    \label{fig:collection_distributions}
\end{figure}

The resulting spatial collection distribution is shown in \autoref{fig:collection_distributions}. The calculation assumes SPADs with a diameter of 12~\textmu m (matching the chip design in \autoref{sec: CMOS SPAD Array Design}), a 10~\textmu m thick diamond layer, and a 10~\textmu m air gap between the diamond and the SPAD array.

To compare different designs, we define several performance metrics. The \emph{peak collection efficiency} is the collection efficiency for an NV center located directly above the center of the SPAD. The spatial profile of the collection function is also important, as it determines sensitivity to crosstalk and spatial resolution. Finally, we define the \emph{effective collection surface} (ECS) as the total collected fluorescence normalized by the NV concentration, yielding a quantity with units of \textmu m$^2$. This corresponds to the equivalent diamond area that would emit the same number of photons into the detector.

Numerical evaluation shows that the simple design has a peak collection efficiency of approximately 0.75\% (see \autoref{fig:collection_distributions}). Its broad collection profile makes it more susceptible to optical crosstalk. For the present chip with a 55~\textmu m pixel pitch, this effect is negligible, but future designs with smaller pitch may experience significant crosstalk. The dependence on pixel pitch is shown in \autoref{fig:optical_crosstalk}. The ECS of this design is 4.6~\textmu m$^2$.

\subsubsection*{Quartz Micropillars}
\begin{figure}
    \centering
    \includegraphics[width=\linewidth]{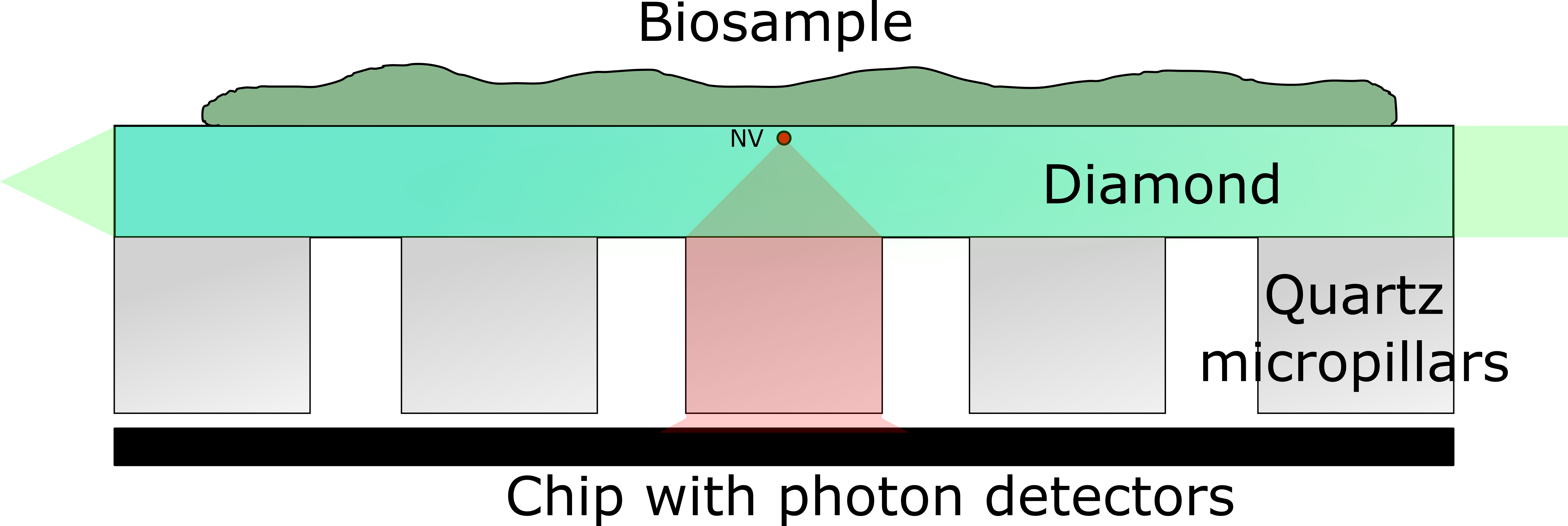}
    \caption{The quartz micropillar design. For a 10 \textmu m thick diamond, it has a peak collection efficiency of 4.7\%, which is significantly higher than the other two designs. Furthermore, it has a narrow spatial profile. That makes it a good candidate in circumstances where inhomogeneous ODMR broadening is a problem at the scale of tens of microns. It collects the equivalent of 5.2 \textmu m $^2$ of diamond, making the total photon count just marginally better than the simple design.}
    \label{fig:design_quartz}
\end{figure}

In this design, the collection cone is defined by total internal reflection at the diamond–quartz and quartz–air interfaces, with critical angles of approximately $37^\circ$ and $43^\circ$, respectively (see \autoref{fig:design_quartz}). The coupling is determined by the overlap between the NV emission cone and the aperture of the quartz pillar.

For the chosen values above, this geometry yields a peak collection efficiency of 4.7\% and a sharp spatial cutoff for NV centers displaced from the pillar axis (see \autoref{fig:collection_distributions}). The ECS is 5.2~\textmu m$^2$, only slightly larger than that of the simple design, but with significantly improved spatial selectivity. The sharp cutoff helps reduce inhomogeneous broadening of ODMR signals.

Optical crosstalk is strongly suppressed because efficient coupling occurs only when the emission cone directly overlaps the pillar entrance. An upper bound of $1.6 \times 10^{-6}$ was obtained for fluorescence entering through the pillar sidewalls, which is therefore negligible.

To avoid overlap of collection regions between adjacent pillars, the minimum center-to-center spacing equals the diameter of the non–TIR-limited emission cone plus the pillar diameter. The pillar diameter must not exceed the SPAD active area to avoid light loss. Choosing the pillar diameter equal to the SPAD diameter maximizes collection. For a 10~\textmu m thick diamond layer and a diamond–quartz critical angle of $37^\circ$, the minimum spacing is 27~\textmu m, well below the 55~\textmu m pixel pitch of the present chip. Consequently, no measurable crosstalk is expected. Pixel pitch could be reduced to approximately 27~\textmu m before crosstalk becomes significant.

\subsubsection*{Microlenses}
Diamond microlenses are incompatible with TIR excitation, which requires a planar interface to sustain total internal reflection. However, for single-NV applications using non-TIR excitation, microlenses can efficiently focus emission from individual NV centers onto detectors.

Using ray optics, the maximum achievable collection angle corresponds to a hemispherical lens profile. In this case, light remains collimated provided that the incidence angle at the lens surface remains below the TIR threshold. The maximum internal collection angle is therefore
\[
\theta_{\mathrm{col,max}} = 90^\circ - \theta_{\mathrm{TIR}} = 65.6^\circ,
\]
corresponding to approximately 29\% of the emitted fluorescence being collected for a centrally located NV center. The efficiency decreases rapidly for off-axis emitters, resulting in a small effective collection volume. Nevertheless, this performance exceeds that of the other designs considered here for non-TIR excitation schemes.

Quartz microlenses can be used in TIR-compatible configurations, but the lower refractive index limits the collection angle to approximately $26^\circ$, corresponding to about 5.1\% collection efficiency.

\subsubsection*{The Diamond-Quartz-Diamond design}
\begin{figure}
    \centering
    \includegraphics[width=\linewidth]{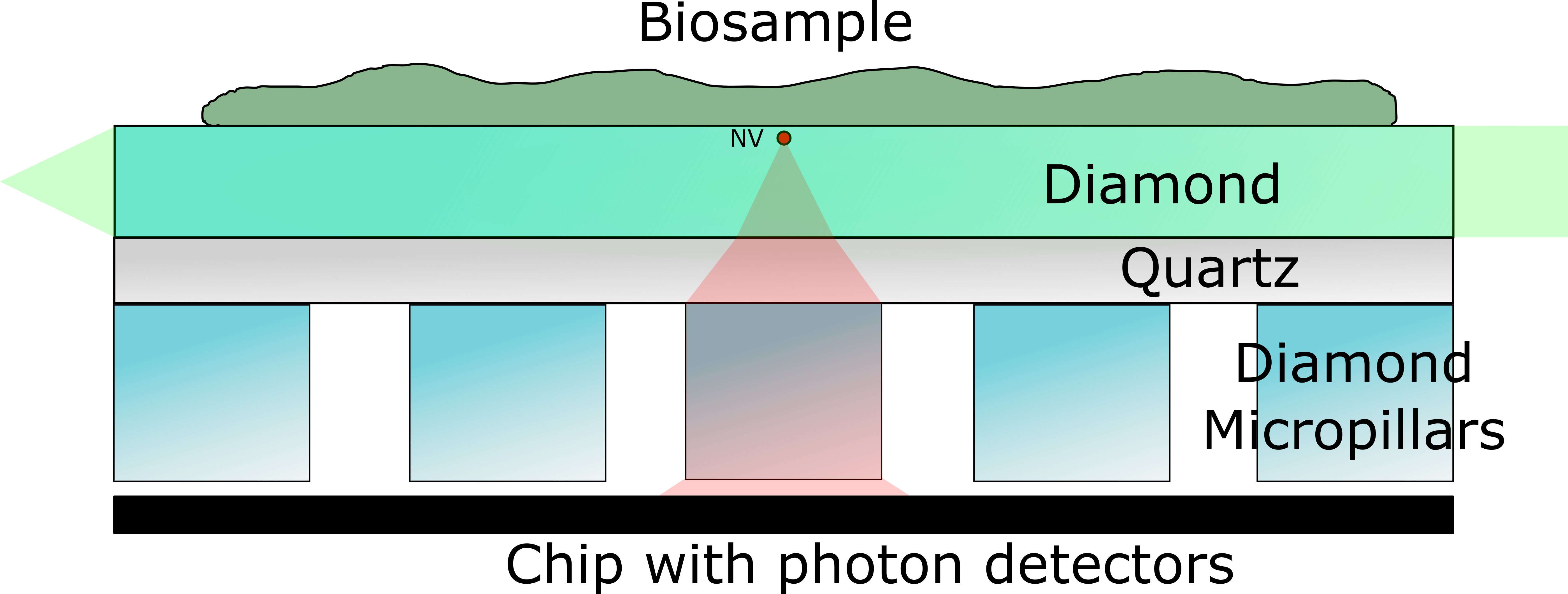}
    \caption{The diamond-quartz-diamond design. This is a more complex design that aims to keep the excitation light internally reflected, while collecting the most fluorescence light possible by geometric optics. It has the same profile as the simple design. For 10 \textmu m thick diamond and quartz layers, it has a peak collection efficiency of 1.5\%. Nonetheless, it collects the equivalent of 11.4 \textmu m $^2$ of diamond, more than double that of the other two designs.}
    \label{fig:design_diamond_pillars}
\end{figure}

A planar NV-containing diamond slab mounted on quartz, with diamond micropillars etched below the quartz layer, can redirect fluorescence toward the SPAD array (see \autoref{fig:design_diamond_pillars}). This configuration maintains TIR excitation while enhancing collection.

\begin{figure}
    \centering
    \includegraphics[width=\linewidth]{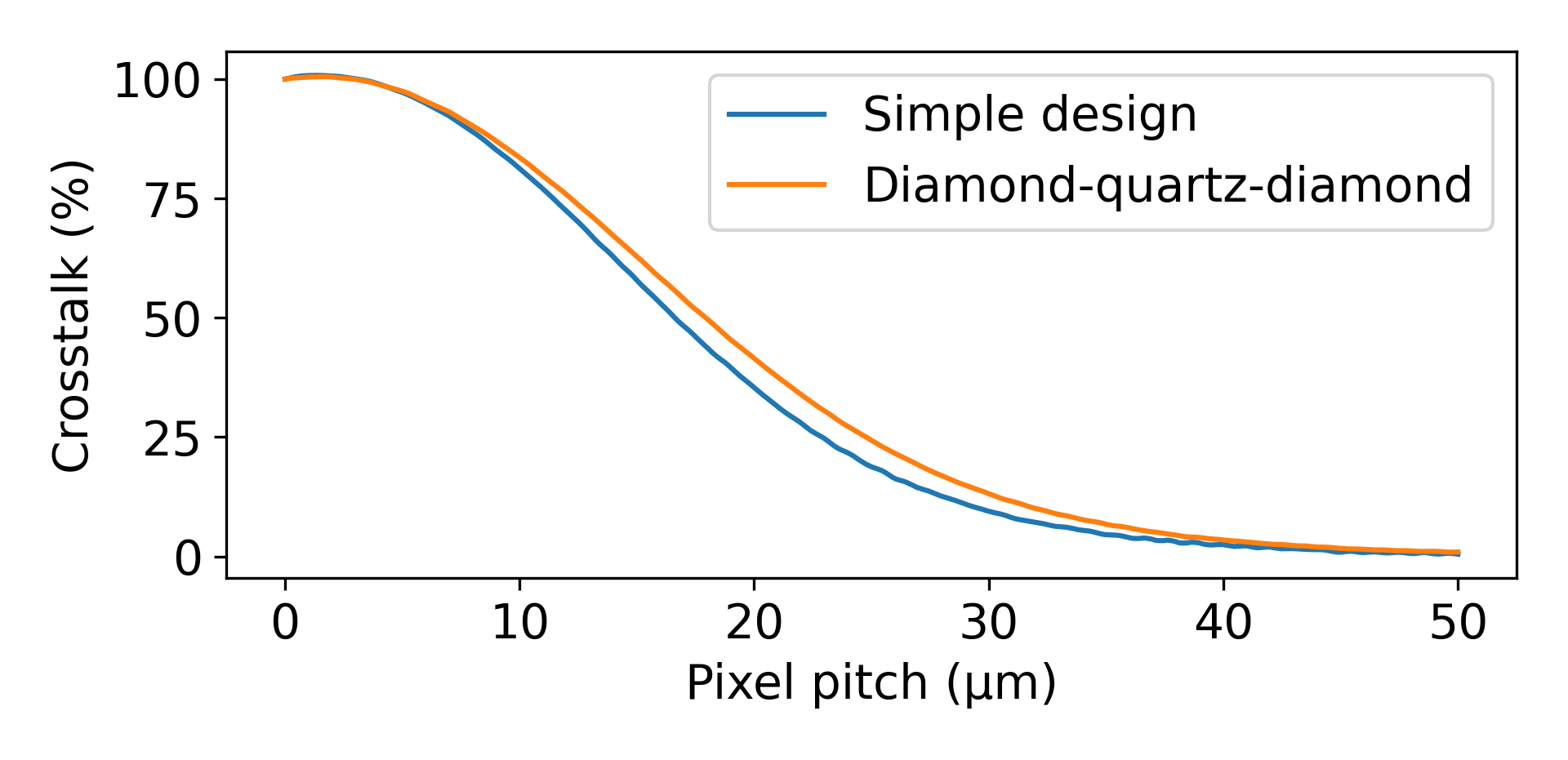}
    \caption{The optical crosstalk between pixels vs pixel pitch. Some NV centers will reach two or more pixels simultaneously, resulting in optical crosstalk on top of any electrical crosstalk that might be present in the SPAD chip. The quartz pillar design is excluded from this graph, since its crosstalk was found to be negligible. However, note that this holds true only if pixel pitch of the quartz pillars is at least 27 \textmu m, which softly coincides with the minimum distances from the other two designs. The two graphs differ slightly due to having different Fresnel coefficients. The chip presented in \autoref{sec: CMOS SPAD Array Design} has a pixel pitch of 55 \textmu m and will thus have negligible crosstalk.}
    \label{fig:optical_crosstalk}
\end{figure}

Simulations assuming 10~\textmu m thick diamond and quartz layers yield a peak collection efficiency of 1.5\% and an ECS of 11.4~\textmu m$^2$, more than double that of the other designs despite the lower peak efficiency. Sidewall coupling into the pillars is neglected as in the quartz-pillar analysis. Crosstalk can still occur via propagation within the quartz layer, producing a spatial profile similar to that of the simple design (see \autoref{fig:optical_crosstalk}).

\subsubsection*{Objective lens (NA~0.95) for comparison}
An objective with NA = 0.95 corresponds to a half-angle of $71.8^\circ$ in air, which refracts to approximately $23.1^\circ$ inside diamond according to Snell’s law. This corresponds to a geometric collection efficiency of roughly 4\%.

\subsection{Our design choice}
\textbf{Single NV vs ensemble NV}
Single NV and ensemble NV diamonds both come with their advantages and disadvantages, depending on the targeted application. Using NV ensemble allows for a much higher signal (and thus better signal-to-noise ratio). This is ultimately the main reason of ensemble's success, despite the worse coherence times and inhomogeneous broadening. Ensembles are also easier to fabricate and integrate in a sensor. Single NV diamonds have excellent spatial resolution due to the NV center's atomic sizes and isolated operation. This gives single NV the upper hand when extremely small spatial resolution is needed or when coupled with a micro-magnet. For the compact bio-sensor, we suggest using ensemble NV diamonds.

\noindent
The simplistic diamond-slab-on-chip design requires less effort and tools to fabricate compared to the other discussed designs. Nonetheless, each such pixel collects the equivalent light of 4.6 \textmu m $^2$ of diamond. That is enough to saturate the SPADs, as will become clear in \autoref{sec:sensitivity}. Hence, the rest of this paper will focus on the diamond-slab-on-chip design.

\section{Efficiency}
\label{sec:efficiency}

The overall performance of the optical subsystem depends not only on the chosen excitation and collection geometry but also on the efficiency of light transmission through the stack and the detection probability at the SPAD array. Building on the concepts outlined above, the system-level efficiency can be decomposed into four main contributions:

\subsubsection*{Geometric efficiency} As discussed in \autoref{sec:excitation_and_collection}, most optical stack designs collect only a fraction of the NV emission cone. In objective-based schemes this is limited by the numerical aperture, while in TIR- or waveguide-based schemes it is defined by critical angles at the diamond interfaces. Lower-index surroundings reduce the in-diamond acceptance cone and therefore collection efficiency, though careful index engineering (e.g., using quartz layers or micropillar structures) can mitigate this limitation.

\subsubsection*{Transmission efficiency} 
\begin{figure}
  \centering
  \includegraphics[width=\linewidth]{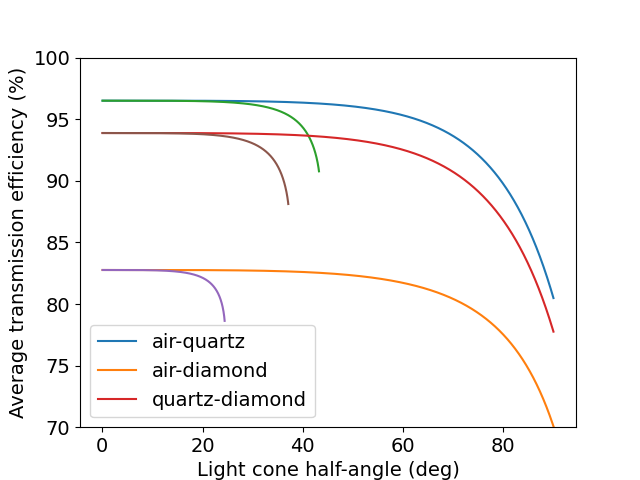}
  \caption{Average transmission efficiency for three relevant materials. The transmission is calculated assuming unpolarized light (equal amounts of s- and p-waves) and averaged over the full solid angle. The shorter curves represent the same surface in reverse. Note that swapping the surface does not affect the transmission efficiency for small angles. The swapped-surface curves stop at lower angles due to reaching total internal reflection.}
  \label{fig:fresnel_transmission}
\end{figure}

Fresnel transmission at each interface (diamond–quartz–air) depends on incidence angle and polarization. For NV ensembles, which emit with mixed dipoles and polarizations, the average transmission of an unpolarized point source approximates the emission. By integrating the Fresnel transmission formula over different angles, a graph is obtained of average transmission efficiency vs light cone half-angle, see \autoref{fig:fresnel_transmission}. For a wide range of angles, the average transmission efficiency is well-approximated by the perpendicular transmission coefficient. This figure is included to give the reader an intuitive feeling of to what extent Fresnel losses matter.

\subsubsection*{Detector efficiency} The final contribution is the photon detection efficiency (PDE) of the SPADs, which must be evaluated across the NV fluorescence band (600–800~nm). The effective system quantum efficiency also depends on the integrated metal-grating filter and any additional optical stack elements that modify transmission. Typical CMOS SPADs have a PDE around 15\% for 600 - 800 nm, for example in M. W. Fishburn \cite{fishburnFundamentalsCMOSSinglephoton2012} and in S. Pellegrini et al \cite{pellegriniIndustrialisedSPAD402017}. However, special red-enhanced SPADs can be made. Takai et al \cite{takaiSinglePhotonAvalancheDiode2016}] made a CMOS-compatible red-enhanced SPAD that has a PDE of 60\% for 600 nm and 20\% for 870 nm at room temperature. For the typical NV emission, which consists mostly of the sideband, this would average to about 40\% efficiency. Regarding metal gratings, \cite{ibrahimHighScalabilityCMOSQuantum2021} reports -2.1 dB for red (700 nm) and -21.8 dB for green (532 nm) light for their metal grating, on top of -3 dB due to the grating covering half of the SPAD surface. This provides an excellent option in the case there is a problem of green light leakage to the SPAD.

\subsubsection*{Excitation coupling} For TIR-based excitation, side coupling via free space or fiber can inject nearly the full pump beam into the diamond. Fresnel reflections at the air–diamond facet are minimal for near-normal incidence but increase for larger angular spectra, with surface roughness introducing additional scattering losses. The use of index-matching gels can reduce Fresnel reflections \cite{indexgel}. A simple lower bound for the transmission with index matching gel can be obtained by assuming the gel has the same refractive index as the optical fiber's index. Then, the only loss will be at the gel-diamond interface. Since the gel's refractive index is similar to quartz in this simplified case, the transmission of quartz-diamond interface is representative, which is given in \autoref{fig:fresnel_transmission}. For a wide range of angles, the transmission is 94\%.

\subsubsection*{Errors in alignment} TIR-based designs with NV ensembles are very robust for small misalignments, as no particular NV center has to be in focus and excitation light is delivered as a flooding wave that is practically indifferent to small translations and rotations.

\section{Sensitivity estimation of the diamond slab on chip TIR design}
\label{sec:sensitivity}
In this subsection, an estimation is made of the magnetic field sensitivity of the diamond-slab-on-chip design, also referred to as the "simple design". The light is collected by the SPADs without any guidance. Nevertheless, the SPADs will detect sufficient fluorescence photons, which is explained later in this subsection. The exact number of counts depends on the NV density in the NV layer of the diamond.

\subsubsection*{Total efficiency of fluorescence detection}
In order for the fluorescence photons to be counted, they have to traverse multiple steps, which can be categorized as geometric collection efficiency, transmission efficiency and detector efficiency.

For the simple design, recall that its equivalent collection surface is 4.6 \textmu m $^2$, see \autoref{sec:excitation_and_collection}). This number already includes Fresnel loss at the diamond-air interface.

Finally, the SPADs themselves have photon detection efficiencies (PDE). For the sake of estimating the characteristics of the device described in this paper, we assume the total internal reflection filters enough of the green light such that we may use SPADs without metal grating. This allows usage of bare SPADs, which will have PDEs around 40\% as discussed in \autoref{sec:efficiency}.

\subsubsection*{The sensitivity equation}
\label{subsec:sensitivity_equation}
The following sensitivity calculation will be for Continuous Wave Optically Detected Magnetic Resonance (CW-ODMR), which is the simplest method of ODMR and does not require precicely timed hardware.
Given the full-width half-maximum (FWHM) $\Delta\nu$, the ODMR contrast $C$ and the photon countrate $R$, the magnetic field sensitivity can be estimated by \cite{barrySensitivityOptimizationNVDiamond2019}

\begin{equation}
    \eta_B = \frac{4}{3\sqrt{3}} \frac{1}{\bar{\gamma}_{NV}} \frac{\Delta\nu}{C\sqrt{R}}
    \quad ,
    \label{eq:sensitivity_main}
\end{equation}
\noindent
where $\bar{\gamma}_{NV}$ = 28 MHz/mT is the reduced gyromagnetic ratio of NV centers, defined by \cite{qnamiFundamentalsMagneticField2020}

\begin{equation}
    \bar{\gamma}_{NV} = \frac{g_e \mu_B}{h}
    \quad .
\end{equation}
The constant factor from \autoref{eq:sensitivity_main} comes from the maximum slope present in the corresponding Lorentzian curve.

The countrate $R$ is the rate of photons detected at the detector. It is determined by the number of fluorescence photons in the collection volume and any losses due to optics and photon detection probability (PDP). The number of NV centers $N_{NV}$ can be expressed in terms of NV density $\rho_{NV}$, collection volume $V$ and the fraction of NV centers in the orientation adressed by our bias field $f_{orient}$, yielding the equation

\begin{equation}
    N_{NV} = \rho_{NV} V f_{orient}
\end{equation}

Let $r_{NV}(s)$ be the rate of fluorescence photons emitted per NV center for a given saturation $s = P_{opt}/P_{sat}$. Here, $P_{opt}$ is the power of the optical pumping, and $P_{sat}$ is the corresponding saturation power. The countrate of the detector is expressed in terms of $r_{NV}$ by

\begin{equation}
    R = \eta_{optics} \eta_{PDP} N_{NV} r_{NV}(s)
    ,
\end{equation}
or alternatively in terms of NV density $\rho_{NV}$:

\begin{equation}
    R = \eta_{optics} \eta_{PDP} \rho_{NV} V f_{orient} r_{NV} (s)
    .
    \label{eq:countrate}
\end{equation}

The function $r_{NV}(s)$ can be expressed as

\begin{equation}
    r_{NV}(s) = r_{NV}^\infty \frac{s}{1+s}
\end{equation}

The linewidth and contrast depend on the laser and microwave power used for ODMR. Having too much power of both results a power broadened linewidth, reducing sensitivity. Besides that, the laser and microwave powers need to have a balance in order to have the best (lowest) $\Delta\nu / C$ ratio. Besides power broadening, the linewidth is limited by $T_2^*$ by $\Delta\nu \approx \frac{1}{\pi T_2^*}$ \cite{levinePrinciplesTechniquesQuantum2019}, which is important for deciding the best nitrogen concentration for the diamond. The choice of nitrogen concentration is elaborated in \autoref{subsubsec:diamond_properties}. To optimize the laser and microwave powers, the results are used from Dréau et al \cite{dreauAvoidingPowerBroadening2011}, which gives the following formulas:

\begin{equation}
    C = \frac{1}{2} \cdot \frac{(\alpha - \beta) \Gamma_p}{(\alpha + \beta) \Gamma_1 + \alpha \Gamma_p} \cdot \frac{\Omega_R^2}{\Omega_R^2 + \Gamma_2 (2 \Gamma_1 + \Gamma_p)}
    \label{eq:contrast}
\end{equation}

\begin{equation}
    \Delta\nu = \frac{1}{2 \pi} \sqrt{\Gamma_2^2 + \frac{\Omega_R^2 \Gamma_2}{2 \Gamma_1 + \Gamma_p}}
    \label{eq:linewidth}
\end{equation}
In the above formulas, $\alpha = 1$ and $\beta = 0.8$ are the constants that represent the brightness of the spin states. $\Omega_R$ is the Rabi frequency, which scales with microwave amplitude and is typically between 0 and 1 MHz. $\Gamma_1 \approx 10^3 ~\text{s}^{-1}$ is the spin-lattice relaxation rate, and $\Gamma_2^*$ is the inhomogeneous dephasing rate. $\Gamma_p = \Gamma_p^\infty \frac{s}{1+s}$ is the optically induced spin polarization rate, where $\Gamma_p^\infty = 5 \cdot 10^6 ~s^{-1}$. The relaxation rate of the spin coherences is defined by $\Gamma_c = \Gamma_c^\infty \frac{s}{1+s}$, where $\Gamma_c^\infty = 8 \cdot 10^7 ~s^{-1}$. $\Gamma_c^\infty$ is what sets the upper limit to the number of photons a single NV center can emit per second. Finally, $\Gamma_2 = \Gamma_2^* + \Gamma_c$. All above constants are included in \autoref{tab:diamond_and_constants}, which summarizes all relevant constants and values for ODMR.

\subsubsection*{Diamond properties and constants}
\label{subsubsec:diamond_properties}
In order to optimize magnetic sensitivity, the right nitrogen concentration $[N]$ needs to be chosen. In the regime of dense diamonds where the nitrogen spin bath dominates decoherence, increasing $[N]$ will both increase signal and linewidth, such that ultimately the magnetic sensitivity becomes worse \cite{levinePrinciplesTechniquesQuantum2019}. As natural $^{13}\text{C}$ spins limit $T_2^*$, the diamond should be isotopically controlled $^{12}\text{C}$ samples. For such diamonds, other less prominent limiting factors to $T_2^*$ are significant for approximately $[N] < 1 \text{ppm}$ \cite{bauchUltralongDephasingTimes2018}. Hence, $[N] = 1 \text{ ppm}$ is a great choice in order to maximize signal while negligably impacting $T_2^*$. At this concentration, $T_2^* \approx 10 \text{ } \mu\text{s}$ \cite{bauchDecoherenceDipolarSpin2020}.

Despite that, the optimal concentration is actually constrained by the SPAD saturation limit. Too many NV centers will saturate the SPAD, effectively reducing contrast. The best signal-to-noise is achieved by maximizing signal within the limits of the SPAD. Due to the Poissonian nature of shot-noise, the SNR scales as $r^{1/2}$, with $r$ the photon rate. The SPADs of the chip introduced in \autoref{sec: CMOS SPAD Array Design} are designed for 3 Mcps. Hence, around 1 Mcps is the sweet spot for optimal sensitivity. Optimal ODMR sensitivity is typically reached with the saturation parameter $s \sim 10^{-3}$. If the fluorescence of 500 NV centers is collected into each SPAD (or an equivalent amount, such as half cones of 1000 NV centers), then the detected number of counts reaches 1 Mcps at a saturation of $s = 7 \cdot 10^{-3}$. \autoref{fig:sensitivity_simulated} indicates the optimal sensitivity is slightly lower than that, making 500 NV centers per SPAD an ideal amount to work with. The corresponding NV surface density is 500 / $A_{SPAD}$ = 38 $\mu\text{m}^{-2}$ = $3.8 \cdot 10^9 ~\text{cm}^{-2}$. That corresponds to NV density in the order of ppb, and thus a nitrogen density in the order of tens to hundreds of ppb before activation. Note that, at 1 Mcps, any dark counts from the SPAD are negligible, and shot-noise is the dominating noise.

\begin{table}[]
    \centering
    \begin{tabular}{c|c}
        Number of NV centers per SPAD & 500 \\
        NV surface density & $3.8 \cdot 10^9$ cm$^{-2}$ \\
        Spin-lattice relaxation rate $\Gamma_1$ & $10^3 \text{ s}^{-1}$ \\
        Saturated polarization rate $\Gamma_p^\infty$ & $5 \cdot 10^6 \text{ s}^{-1}$ \\
        Spin relaxation optical cycle rate $\Gamma_c^\infty$ & $8 \cdot 10^7 \text{ s}^{-1}$ \\
        Fraction of NV in measurement orientation $f_{orient}$ & $\frac{1}{4}$ \\
        Maximum contrast $\Theta$ & 0.2
    \end{tabular}
    \caption{Relevant diamond properties and other constants.}
    \label{tab:diamond_and_constants}
\end{table}

\subsubsection*{Optimizing optical pumping power}
\begin{figure}
    \centering
    \includegraphics[width=1\linewidth]{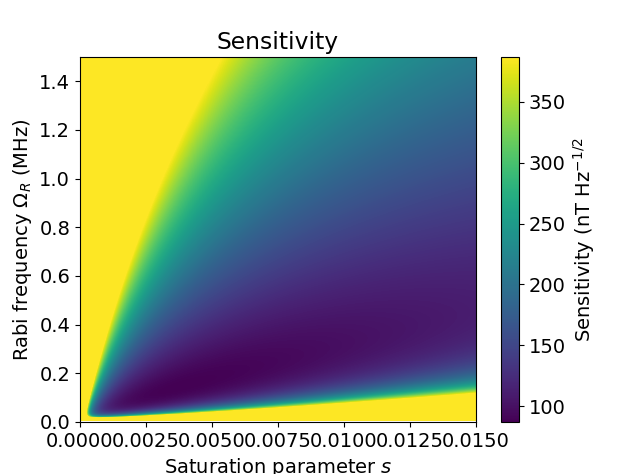}
    \caption{Simulated sensitivity plot for a range of Rabi frequencies and saturation values. The calculation takes into account all efficiencies of the proposed device. The optimal (minimum) sensitivity is approximately 90 nT/$\sqrt{\text{Hz}}$.}
    \label{fig:sensitivity_simulated}
\end{figure}

Recall that the magnetic field sensitivity $n_B$ is determined by \autoref{eq:sensitivity_main}. Substitute $\Delta\nu$, $C$ and $R$ with their given formulas \ref{eq:linewidth}, \ref{eq:contrast}, \ref{eq:countrate} respectively. This yields an equation that relates magnetic field sensitivity to Rabi frequency $\Omega_R$ and saturation parameter $s$, which can then be converted to real microwave power and laser power that have to be used in the device. A 2D sweep is plotted, which is shown in \autoref{fig:sensitivity_simulated}. The minimum achievable sensitivity is approximately 90 nT/$\sqrt{\text{Hz}}$, with a corresponding linewidth of 0.15 MHz. Future realisations of this concept might require higher NV densities for optimal sensitivity, if the SPADs are made smaller to accomodate more pixels per surface area. However, as mentioned above, decoherence by nitrogen interactions does not start dominating $T_2^*$ until [N] $\sim$ 1 ppm, which is about one to two orders of magnitude larger than the density proposed above. Hence, enough margin is available for future realizations of this concept sensor.

\subsection{Delivery of the excitation light}
In order to excite the NV centers in the diamond, the excitation light has to enter the diamond first. On top of that, the light has to be uniform, in order to be able to operate the NV centers at the most sensitive laser and microwave intensities. A possible modular and portable design is guiding excitation light through an optical fiber into the side of the diamond. Let us make the following assumptions: The fiber is edge-coupled to the diamond. The outgoing beam has a Gaussian profile. By the many different angles and reflections happening, the light is assumed to become decoherent. The Gaussian beam is approximated as coming from a point source, which is valid for large distances $\gg$ $w_0$, with $w_0$ the waist size of the beam. The light reaching the opposite end of the diamond is disregarded. In practice, part of this light escapes the diamond while a part reflects and will keep internally reflecting until it reaches the starting side again, etcetera.

\begin{figure}
    \centering
    \includegraphics[width=\linewidth]{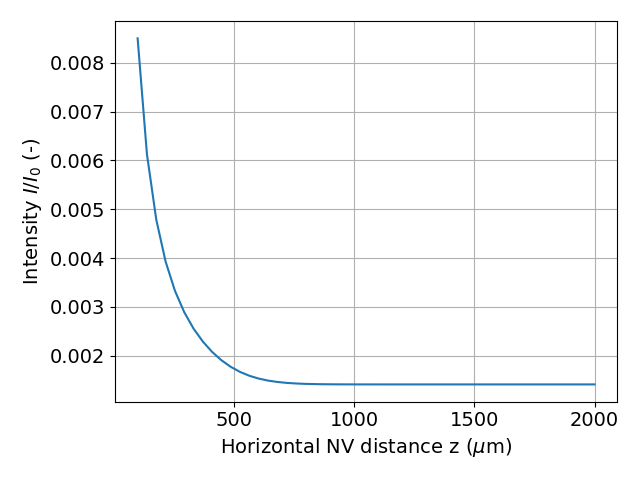}
    \caption{Excitation light intensity in the NV layer at different distances from the side of the diamond. In the first 750 micron, the laser spreads out by the first total internal reflections. Afterwards, the laser is practically fully spread out and stays uniform for arbitrarily long distances. This simulation is for a diamond with thickness $10 \mu$m, width 1 mm, the NV layer 20 nm below the diamond surface, and with a divergence angle 53$^\circ$, which occupies the full TIR angle.}
    \label{fig:field_intensity}
\end{figure}

In order to estimate the light intensity in the NV layer, the method of images is used. The many internal reflections are equivalent to an infinite plane of identical beam sources. The beam is modeled as a Gaussian beam, given by the formula

\begin{equation}
    I(r, z) = I_0 \left( \frac{w_0}{w(z)} \right)^2 \exp \left( \frac{-2 r^2}{w(z)^2} \right)
    ,
\end{equation}
where $I(r, z)$ is the intensity at a radial distance $r$ from the beam and an axial distance $z$, $I_0$ is the intensity at the waist, $w_0$ is the waist size of the beam, $w(z) = z \tan(\theta_{div})$ is the approximate 1/e$^2$ threshold of the beam at a given $z$ and divergence angle $\theta_{div}$. The method of images is then applied by constructing an infinite series:

\begin{align}
    r_{m,n} =& \sqrt{(h + n d)^2 + (m w)^2} \\
    I_{tot}(z) =& \sum_{m,n} I(r_{m,n}, z)
    \quad .
\end{align}
Here, $m$ and $n$ are integers that iterate the mirror images in the direction of the width and height of the diamond respectively. $d$ is the thickness of the diamond, $w$ is the width of the diamond, and $r_{m,n}$ is the radial distance between the beam and the NV center. $h$ is an extra offset in height of the NV center, to account for it being near the diamond surface and not necessarily in the middle of a beam. Due to the beam being confined within a cone of its divergence angle, the series can be truncated to a finite series that is trivially computed. The maximum possible divergence angle that can be used is limited by the angle of total internal reflection between the diamond and the biosample. The biosample consists of mostly water, and will thus have a refraction index of approximately 1.5, which corresponds to a TIR angle of 37$^\circ$. Hence, $\theta_{div,max} = 90^\circ - 37^\circ = 53^\circ$. That is orders of magnitude broader than the typical outgoing beam from a single mode fiber, and still multiple times larger than commonly seen for multi-mode fibers. However, with the help of diffusers, angles can be more spread out. \autoref{fig:field_intensity} shows the intensity at the NV layer for a Gaussian beam that occupies the full range of possible TIR angles. After about 750 micron, it converges to a nonzero value and stays uniform. This uniform region is then suitable for NV sensing. This indicates that a 1.75x1.0 mm$^2$ diamond would be suitable for having a 1x1 mm$^2$ sensing area, where 0.75 mm is used for spreading the excitation light.

\section{Device proposal, outline and characteristics}
In this section, the overall system and characteristics are outlined. The chip design presented in \autoref{sec: CMOS SPAD Array Design} is only for a proof-of-concept. Future iterations with more pixels and a smaller pixel pitch are a must. The device has a field-of-view of 880x880 $\mu\text{m}^2$. Its magnetic field sensitivity is 90 nT/$\sqrt{\text{Hz}}$ (see \autoref{subsec:sensitivity_equation}). The pixel pitch is 55 $\mu\text{m}$. The TIR and collection work at any size, but SPADs are limited to tens of microns. The physical pixels (including the diamond micropillars) have to match the SPAD pixels. Hence, the pixel pitch of the overall system is limited to the pitch of the CMOS chip. A smaller pixel pitch is desirable for more precise sensing. This can be achieved with denser design, including 3D CMOS design. However, this comes at a tradeoff with cross talk between SPAD pixels.

The chip's pixel pitch of 55 $\mu\text{m}$ is not enough for cellular precision cancer detection, but it is still sufficient for cardiography, for which millimeter-scale sensing is required \cite{araiMillimetrescaleMagnetocardiographyLiving2022}. A future iteration of the chip may be designed with a smaller pixel pitch sufficient for sensing at single-cell precision.

The thickness of the diamond may be between 0 and 120 micron. The other dimensions of the diamond are 1.75 mm long and 1.0 mm wide. 0.75 mm of its length is reserved for spreading the injected excitation light. For other chips with different pixel pitches, the maximum diamond thickness scales linearly with pixel pitch.

The system described is just the chip head. It has to be hooked up to an excitation laser, a suitable microwave source, and electronics to handle the readout of the chip and make the data available on a computer. Most of these devices can also be miniaturized, but that is outside the scope of this paper.

The diamond and the SPAD array do not require a particular alignment. Hence, they can be glued together with any glue that is optically clear and does not electronically interact with the CMOS chip. Since the refraction index of any common glue is between the index of diamond and air, it will also serve as index-matching gel and thus improve losses due to internal Fresnel reflections. For excitation light delivery, an optical fiber has to be connected to the side of a diamond. Due to the small (micron-) sizes involved, succesfully connecting the fiber is a challenge. This may require a small alignment construction. Alternatively, light can be focused on the side of the diamond in freespace for applications that rely less on portability.


\section{Biosensing}
\label{sec:biosensing}

Having established the hardware subsystems for excitation, fluorescence collection, MW delivery, and digital readout, we now turn to the biological use case that motivates this integration effort. Biosensing with optically detected magnetic resonance (ODMR) has emerged as a powerful approach for probing the subtle magnetic environments generated by biological samples. HEK293T cells are used here as a model system: when labeled with superparamagnetic iron oxide nanoparticles (SPIONs), which magnetize under an external bias, the resulting magnetic perturbations can be quantitatively imaged with NV centers in diamond. This strategy provides a window into the biophysical properties of cells and their immediate microenvironment, making it highly relevant for both fundamental research and diagnostic applications.

The integration of SPIONs with HEK293T cells is central to this biosensing approach. HEK293T cells, known for their robust growth and ease of genetic manipulation, are incubated with biocompatible \ce{Fe3O4} nanoparticles. Due to their superparamagnetic behavior, these nanoparticles acquire a magnetic moment when exposed to an external field. Although in the present work the SPIONs were not functionalized to target specific biomarkers, their random dispersion on the cell surface is sufficient to demonstrate proof-of-principle magnetic imaging. This labeling approach enables the detection of even subtle differences in cellular states, thereby allowing discrimination between physiological conditions.  

\begin{figure}
    \centering
    \includegraphics[width=\linewidth]{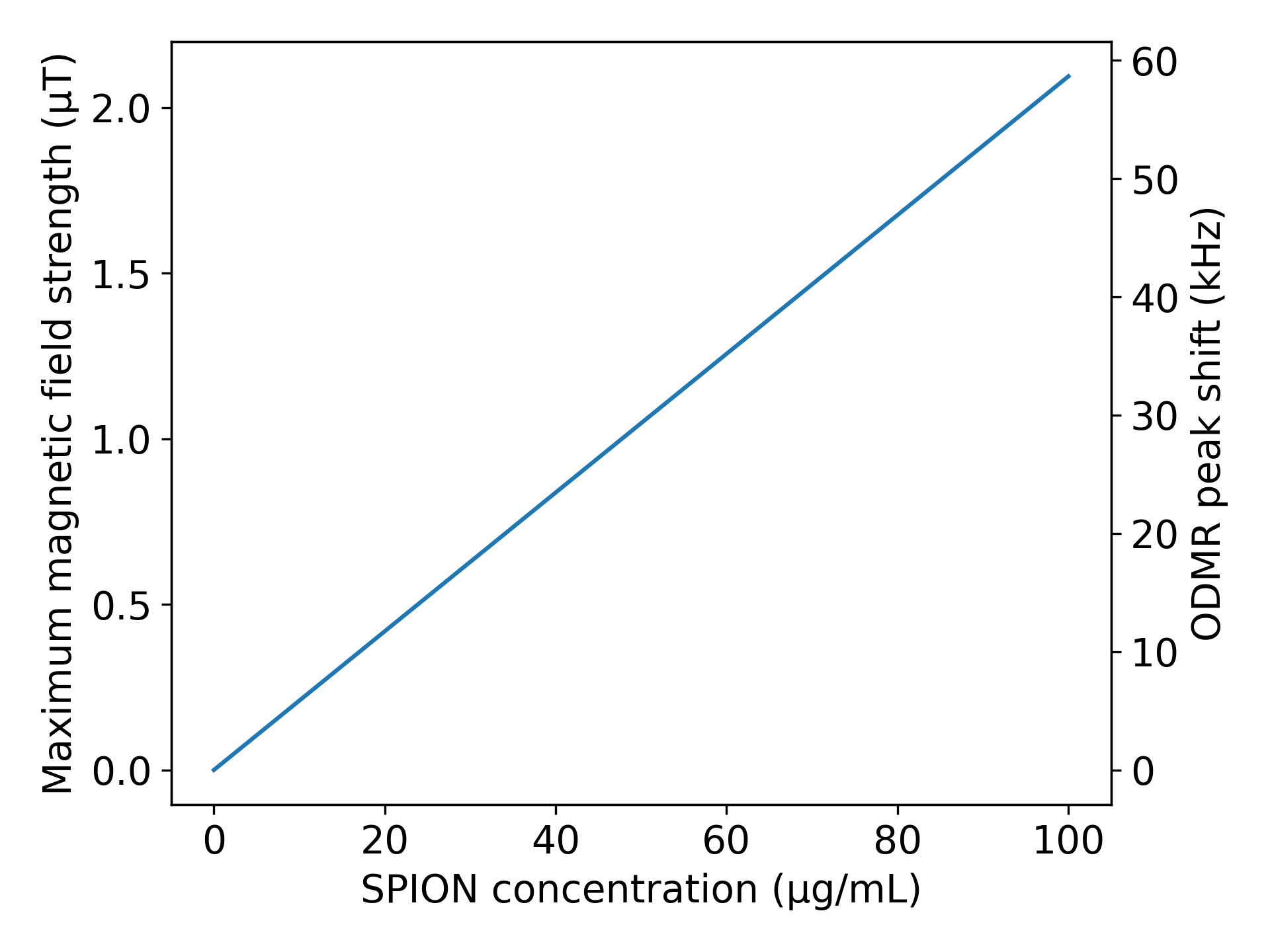}
    \caption{Maximum magnetic field strength at the surface of magnetized cells, as a function of SPION concentration. The corresponding ODMR peak shift is shown on the right scale. The cell-NV distance is assumed to be negligibly smaller than the radius of the cell. Note that the real magnetic field observed by the NV center might be lower, down to half of the maximum strength in the worst possible angle.}
    \label{fig:B_vs_c}
\end{figure}

To estimate the strength of SPION-induced fields, we model the HEK293T cell as a uniformly magnetized sphere of radius $R \approx 9~\mu$m (corresponding to a volume of $3 \times 10^{-9}$~mL). With a SPION magnetization per unit mass of $M_m = 25$~emu/g, the effective dipole moment of a labeled cell scales linearly with nanoparticle concentration as
\begin{equation}
    \mathbf{m} = c V M_m,
    \label{eq:dipole_vs_concentration}
\end{equation}
where $c$ is the concentration and $V$ is the cell volume. Substituting typical values yields $m = 3 \times 10^{-12}$~emu ($3 \times 10^{-15}$~A·m$^2$). Using the dipole-field expression,
\begin{equation}
    \frac{\mu_0 m}{4 \pi R^3} < B < \frac{2 \mu_0 m}{4 \pi R^3},
    \label{eq:B_of_cell_m}
\end{equation}
the magnetic field at the cell surface lies between $0.4$ and $0.8~\mu$T for a concentration of $40~\mu$g/mL, a level generally considered safe for viability \cite{weiSuperparamagneticIronOxide2021}. Detecting this dipole at the worst-case orientation corresponds to an ODMR peak splitting of $\sim 0.01$~MHz, which must be resolved against ensemble NV linewidths up to $\sim 0.15$~MHz. This requirement highlights the importance of the system-level optimizations discussed earlier—efficient $B_1$ delivery, high-throughput collection, and low-noise SPAD readout. The relationship between SPION concentration, magnetic field strength, and the resulting ODMR peak shift is plotted in \autoref{fig:B_vs_c}.

While increasing SPION concentration strengthens the signal, it can compromise viability. Toxicity varies with nanoparticle formulation but often emerges above $\sim$100~$\mu$g/mL \cite{weiSuperparamagneticIronOxide2021}. At such levels, fields of several $\mu$T are expected, reinforcing the need for optimized optical throughput and detector performance to remain sensitive at lower, biocompatible concentrations.

An external magnetic field is required to magnetize the SPIONs. To maintain stable magnetization against thermal fluctuations, the dipole energy $E_B = mB$ must exceed the thermal energy scale $k_B T$ at room temperature ($\sim 4 \times 10^{-21}$~J). For example, a 5~nm magnetite nanoparticle with saturation magnetization of 25~emu/g has a dipole moment of $8.5 \times 10^{-21}$~A·m$^2$. In this case, external fields on the order of Teslas would be needed to stabilize magnetization, illustrating why particle size is critical: larger SPIONs possess larger moments, are less susceptible to Brownian fluctuations, and can be magnetized with lower fields. However, larger particles can reduce cellular uptake and increase toxicity, creating a trade-off between size and field strength. By Wei et al., SPION sizes between 10 and 100~nm are considered optimal; in this regime, external fields of a few mT up to several hundred mT are generally sufficient to achieve near-saturation magnetization. 

Once the cells are labeled, ODMR measurements can be used to directly monitor the magnetic fields generated by the SPIONs. The diamond substrate, containing a shallow ensemble of NV centers, is placed in close contact with the cell layer to maximize sensitivity to near-surface magnetic fields. Under green excitation and MW driving, ODMR spectra reveal shifts in the NV spin resonances that correspond directly to the local magnetic environment. These shifts provide a quantitative measure of SPION-induced magnetization and enable the construction of magnetic field maps that reflect cellular activity.

Recent advances in NV surface preparation, SPION synthesis, and ensemble readout techniques are further enhancing both the sensitivity and resolution of such measurements. In our tests, the integration of SPAD-based readout with tailored MW delivery and efficient optical collection demonstrated that sub-microtesla fields, characteristic of biocompatible SPION concentrations, can be detected within ensemble ODMR linewidths. The ability to perform static imaging of HEK293T cells thus validates the feasibility of widefield NV-based biosensing and highlights the practical relevance of our integrated platform.

Looking forward, the transition from proof-of-concept demonstrations to fully integrated, chip-based biosensors represents a critical next step. Such platforms are expected to reduce system complexity and cost, while enabling real-time monitoring of dynamic biological processes under physiologically relevant conditions. By combining quantum sensing methodologies with established biological models, this work charts a pathway toward portable diagnostic tools and deeper insights into cellular function, thereby positioning CMOS-integrated NV biosensors as a promising technology for quantitative, real-time bioimaging.

\section{Conclusion}

\label{sec:conclusion}
We have presented a CMOS-integrated quantum diamond biosensing platform that combines an NV-diamond sensing layer with a custom SPAD array and a compact digital readout chain (FPGA → Arduino → PC). The SPAD array incorporates active quenching and digital hold-off, enabling high count rates while maintaining low noise. Together with closely integrated microwave radiators and practical optical excitation and collection schemes—including objective-based, TIR-based, and waveguide approaches—the system is designed for widefield ODMR with chip-level scalability and reduced experimental complexity.

From a biosensing perspective, the SPION–HEK293T case study defines the expected operating regime for ensemble NV detection. Simple dipole estimates indicate sub-µT magnetic fields at the cell surface for biologically safe SPION concentrations, corresponding to ODMR splittings of roughly 0.01 MHz compared with typical ensemble linewidths of up to ~0.15 MHz (as calculated in \autoref{sec:sensitivity}). These requirements motivate the reported hardware specifications: rapid quench–recharge cycles to support multi-Mcounts/s per pixel, low dark-count electronics, efficient photon collection, and background suppression via the integrated metal-grating filter. Under these conditions, the sensitivity is expected to be limited primarily by photon statistics rather than electronic readout constraints.

Several elements must still be experimentally validated to establish a complete sensitivity budget. These include the end-to-end optical throughput (including interface and Fresnel losses), wavelength-dependent SPAD photon detection efficiency across the NV emission band, interactions between planar TIR excitation and subsurface micro-optics, thermal and magnetic drift during continuous operation, and trade-offs in SPION size and concentration for long-term imaging. Near-term work will therefore focus on calibrated sensitivity measurements, validation on reference phantoms and fixed HEK293T samples with known SPION loading, and co-design of the optical stack with the microwave radiator to achieve uniform bias magnetic fields without degrading fluorescence collection.

Looking ahead, the architecture supports several natural extensions. These include scaling to larger arrays (and potentially 3D integration), implementation of time-gated or pulsed ODMR protocols, FPGA-based lock-in detection for improved contrast, and biocompatible packaging for real-time measurements on living cells. Such developments would further improve sensitivity while maintaining the compact form factor.

Overall, by co-optimizing microwave delivery, excitation and collection geometry, and low-noise single-photon detection around biologically relevant magnetic field levels, this work outlines a practical path from laboratory-scale quantum diamond microscopes toward portable, CMOS-integrated devices capable of quantitative, real-time magnetic imaging in complex biological environments.


\bibliographystyle{ieeetr}
\bibliography{references}

@article{glennSinglecellMagneticImaging2015,
    author = {Glenn, David R and Lee, Kyungheon and Park, Hongkun and Weissleder, Ralph and Yacoby, Amir and Lukin, Mikhail D and Lee, Hakho and Walsworth, Ronald L and Connolly, Colin B},
    doi = {10.1038/nmeth.3449},
    issn = {1548-7091, 1548-7105},
    journal = {Nature Methods},
    langid = {english},
    month = {August},
    number = {8},
    pages = {736--738},
    title = {Single-Cell Magnetic Imaging Using a Quantum Diamond Microscope},
    urldate = {2025-11-04},
    volume = {12},
    year = {2015}
}

@article{hallHighSpatialTemporal2012,
    author = {Hall, L. T. and Beart, G. C. G. and Thomas, E. A. and Simpson, D. A. and McGuinness, L. P. and Cole, J. H. and Manton, J. H. and Scholten, R. E. and Jelezko, F. and Wrachtrup, J{\"o}rg and Petrou, S. and Hollenberg, L. C. L.},
    doi = {10.1038/srep00401},
    issn = {2045-2322},
    journal = {Scientific Reports},
    langid = {english},
    month = {May},
    number = {1},
    pages = {401},
    title = {High Spatial and Temporal Resolution Wide-Field Imaging of Neuron Activity Using Quantum {{NV-diamond}}},
    urldate = {2025-11-04},
    volume = {2},
    year = {2012}
}

@article{mazeNanoscaleMagneticSensing2008,
    author = {Maze, J. R. and Stanwix, P. L. and Hodges, J. S. and Hong, S. and Taylor, J. M. and Cappellaro, P. and Jiang, L. and Dutt, M. V. Gurudev and Togan, E. and Zibrov, A. S. and Yacoby, A. and Walsworth, R. L. and Lukin, M. D.},
    copyright = {2008 Macmillan Publishers Limited. All rights reserved},
    doi = {10.1038/nature07279},
    issn = {1476-4687},
    journal = {Nature},
    langid = {english},
    month = {October},
    number = {7213},
    pages = {644--647},
    publisher = {Nature Publishing Group},
    title = {Nanoscale Magnetic Sensing with an Individual Electronic Spin in Diamond},
    urldate = {2025-12-21},
    volume = {455},
    year = {2008}
}

@article{doldeElectricfieldSensingUsing2011,
    author = {Dolde, F. and Fedder, H. and Doherty, M. W. and N{\"o}bauer, T. and Rempp, F. and Balasubramanian, G. and Wolf, T. and Reinhard, F. and Hollenberg, L. C. L. and Jelezko, F. and Wrachtrup, J.},
    copyright = {2011 Springer Nature Limited},
    doi = {10.1038/nphys1969},
    issn = {1745-2481},
    journal = {Nature Physics},
    langid = {english},
    month = {June},
    number = {6},
    pages = {459--463},
    publisher = {Nature Publishing Group},
    title = {Electric-Field Sensing Using Single Diamond Spins},
    urldate = {2025-12-21},
    volume = {7},
    year = {2011}
}

@article{acostaTemperatureDependenceNitrogenVacancy2010,
    author = {Acosta, V. M. and Bauch, E. and Ledbetter, M. P. and Waxman, A. and Bouchard, L.-S. and Budker, D.},
    doi = {10.1103/PhysRevLett.104.070801},
    journal = {Physical Review Letters},
    month = {February},
    number = {7},
    pages = {070801},
    publisher = {American Physical Society},
    title = {Temperature {{Dependence}} of the {{Nitrogen-Vacancy Magnetic Resonance}} in {{Diamond}}},
    urldate = {2025-12-21},
    volume = {104},
    year = {2010}
}

@article{mochalinPropertiesApplicationsNanodiamonds2012,
    author = {Mochalin, Vadym N. and Shenderova, Olga and Ho, Dean and Gogotsi, Yury},
    copyright = {2011 Springer Nature Limited},
    doi = {10.1038/nnano.2011.209},
    issn = {1748-3395},
    journal = {Nature Nanotechnology},
    langid = {english},
    month = {January},
    number = {1},
    pages = {11--23},
    publisher = {Nature Publishing Group},
    title = {The Properties and Applications of Nanodiamonds},
    urldate = {2025-12-20},
    volume = {7},
    year = {2012}
}

@article{preezElectronParamagneticResonance1965,
    author = {Preez, Du},
    langid = {english},
    title = {Electron {{Paramagnetic Resonance}} and {{Optical Investigations}} of {{Defect Centres}} in {{Diamond}}},
    year = {1965}
}

@article{gruberScanningConfocalOptical1997,
    author = {Gruber, A. and Dr{\"a}benstedt, A. and Tietz, C. and Fleury, L. and Wrachtrup, J. and von Borczyskowski, C.},
    doi = {10.1126/science.276.5321.2012},
    journal = {Science},
    month = {June},
    number = {5321},
    pages = {2012--2014},
    publisher = {American Association for the Advancement of Science},
    title = {Scanning {{Confocal Optical Microscopy}} and {{Magnetic Resonance}} on {{Single Defect Centers}}},
    urldate = {2025-12-20},
    volume = {276},
    year = {1997}
}

@article{maletinskyRobustScanningDiamond2012,
    author = {Maletinsky, P. and Hong, S. and Grinolds, M. S. and Hausmann, B. and Lukin, M. D. and Walsworth, R. L. and Loncar, M. and Yacoby, A.},
    copyright = {2012 Springer Nature Limited},
    doi = {10.1038/nnano.2012.50},
    issn = {1748-3395},
    journal = {Nature Nanotechnology},
    langid = {english},
    month = {May},
    number = {5},
    pages = {320--324},
    publisher = {Nature Publishing Group},
    title = {A Robust Scanning Diamond Sensor for Nanoscale Imaging with Single Nitrogen-Vacancy Centres},
    urldate = {2025-12-20},
    volume = {7},
    year = {2012}
}

@article{maminNanoscaleNuclearMagnetic2013,
    author = {Mamin, H. J. and Kim, M. and Sherwood, M. H. and Rettner, C. T. and Ohno, K. and Awschalom, D. D. and Rugar, D.},
    doi = {10.1126/science.1231540},
    journal = {Science},
    month = {February},
    number = {6119},
    pages = {557--560},
    publisher = {American Association for the Advancement of Science},
    title = {Nanoscale {{Nuclear Magnetic Resonance}} with a {{Nitrogen-Vacancy Spin Sensor}}},
    urldate = {2025-12-20},
    volume = {339},
    year = {2013}
}

@article{balasubramanianNanoscaleImagingMagnetometry2008,
    author = {Balasubramanian, Gopalakrishnan and Chan, I. Y. and Kolesov, Roman and {Al-Hmoud}, Mohannad and Tisler, Julia and Shin, Chang and Kim, Changdong and Wojcik, Aleksander and Hemmer, Philip R. and Krueger, Anke and Hanke, Tobias and Leitenstorfer, Alfred and Bratschitsch, Rudolf and Jelezko, Fedor and Wrachtrup, J{\"o}rg},
    copyright = {2008 Macmillan Publishers Limited. All rights reserved},
    doi = {10.1038/nature07278},
    issn = {1476-4687},
    journal = {Nature},
    langid = {english},
    month = {October},
    number = {7213},
    pages = {648--651},
    publisher = {Nature Publishing Group},
    title = {Nanoscale Imaging Magnetometry with Diamond Spins under Ambient Conditions},
    urldate = {2025-12-20},
    volume = {455},
    year = {2008}
}

@article{rugarProtonMagneticResonance2015,
    author = {Rugar, D. and Mamin, H. J. and Sherwood, M. H. and Kim, M. and Rettner, C. T. and Ohno, K. and Awschalom, D. D.},
    copyright = {2014 Springer Nature Limited},
    doi = {10.1038/nnano.2014.288},
    issn = {1748-3395},
    journal = {Nature Nanotechnology},
    langid = {english},
    month = {February},
    number = {2},
    pages = {120--124},
    publisher = {Nature Publishing Group},
    title = {Proton Magnetic Resonance Imaging Using a Nitrogen--Vacancy Spin Sensor},
    urldate = {2025-12-20},
    volume = {10},
    year = {2015}
}

@article{borettiSingleBiomoleculeImaging2015,
    author = {Boretti, Alberto and Rosa, Lorenzo and Castelletto, Stefania},
    copyright = {\copyright{} 2015 WILEY-VCH Verlag GmbH \& Co. KGaA, Weinheim},
    doi = {10.1002/smll.201500764},
    issn = {1613-6829},
    journal = {Small},
    langid = {english},
    number = {34},
    pages = {4229--4236},
    title = {Towards {{Single Biomolecule Imaging}} via {{Optical Nanoscale Magnetic Resonance Imaging}}},
    urldate = {2025-12-20},
    volume = {11},
    year = {2015}
}

@article{zalieckasQuantumSensingMicroRNAs2024,
    author = {Zalieckas, Justas and Greve, Martin M. and Bellucci, Luca and Sacco, Giuseppe and H{\aa}konsen, Verner and Tozzini, Valentina and Nifos{\`i}, Riccardo},
    copyright = {2024 The Author(s)},
    doi = {10.1038/s42004-024-01182-7},
    issn = {2399-3669},
    journal = {Communications Chemistry},
    langid = {english},
    month = {May},
    number = {1},
    pages = {101},
    publisher = {Nature Publishing Group},
    title = {Quantum Sensing of {{microRNAs}} with Nitrogen-Vacancy Centers in Diamond},
    urldate = {2025-11-04},
    volume = {7},
    year = {2024}
}

@article{balasubramanianUltralongSpinCoherence2009,
    author = {Balasubramanian, Gopalakrishnan and Neumann, Philipp and Twitchen, Daniel and Markham, Matthew and Kolesov, Roman and Mizuochi, Norikazu and Isoya, Junichi and Achard, Jocelyn and Beck, Johannes and Tissler, Julia and Jacques, Vincent and Hemmer, Philip R. and Jelezko, Fedor and Wrachtrup, J{\"o}rg},
    copyright = {2009 Springer Nature Limited},
    doi = {10.1038/nmat2420},
    issn = {1476-4660},
    journal = {Nature Materials},
    langid = {english},
    month = {May},
    number = {5},
    pages = {383--387},
    publisher = {Nature Publishing Group},
    title = {Ultralong Spin Coherence Time in Isotopically Engineered Diamond},
    urldate = {2025-12-20},
    volume = {8},
    year = {2009}
}

@article{webbNanoteslaSensitivityMagnetic2019,
    author = {Webb, James L. and Clement, Joshua D. and Troise, Luca and Ahmadi, Sepehr and Johansen, Gustav Juhl and Huck, Alexander and Andersen, Ulrik L.},
    doi = {10.1063/1.5095241},
    issn = {0003-6951},
    journal = {Applied Physics Letters},
    month = {June},
    number = {23},
    pages = {231103},
    title = {Nanotesla Sensitivity Magnetic Field Sensing Using a Compact Diamond Nitrogen-Vacancy Magnetometer},
    urldate = {2025-12-20},
    volume = {114},
    year = {2019}
}

@article{kuwahataMagnetometerNitrogenvacancyCenter2020,
    author = {Kuwahata, Akihiro and Kitaizumi, Takahiro and Saichi, Kota and Sato, Takumi and Igarashi, Ryuji and Ohshima, Takeshi and Masuyama, Yuta and Iwasaki, Takayuki and Hatano, Mutsuko and Jelezko, Fedor and Kusakabe, Moriaki and Yatsui, Takashi and Sekino, Masaki},
    copyright = {2020 The Author(s)},
    doi = {10.1038/s41598-020-59064-6},
    issn = {2045-2322},
    journal = {Scientific Reports},
    langid = {english},
    month = {February},
    number = {1},
    pages = {2483},
    publisher = {Nature Publishing Group},
    title = {Magnetometer with Nitrogen-Vacancy Center in a Bulk Diamond for Detecting Magnetic Nanoparticles in Biomedical Applications},
    urldate = {2025-12-20},
    volume = {10},
    year = {2020}
}

@article{sturnerIntegratedPortableMagnetometer2021,
    author = {St{\"u}rner, Felix M. and Brenneis, Andreas and Buck, Thomas and Kassel, Julian and R{\"o}lver, Robert and Fuchs, Tino and Savitsky, Anton and Suter, Dieter and Grimmel, Jens and Hengesbach, Stefan and F{\"o}rtsch, Michael and Nakamura, Kazuo and Sumiya, Hitoshi and Onoda, Shinobu and Isoya, Junichi and Jelezko, Fedor},
    copyright = {\copyright{} 2021 The Authors. Advanced Quantum Technologies published by Wiley-VCH GmbH},
    doi = {10.1002/qute.202000111},
    issn = {2511-9044},
    journal = {Advanced Quantum Technologies},
    langid = {english},
    number = {4},
    pages = {2000111},
    title = {Integrated and {{Portable Magnetometer Based}} on {{Nitrogen-Vacancy Ensembles}} in {{Diamond}}},
    urldate = {2025-12-20},
    volume = {4},
    year = {2021}
}

@article{wangPortableDiamondNV2022,
    author = {Wang, Xuemin and Zheng, Doudou and Wang, Xiaocheng and Liu, Xinyu and Qimeng, Wang and Zhao, Junzhi and Guo, Hao and Qin, Li and Tang, Jun and Ma, Zong-Min and Liuc, Jun},
    doi = {10.1109/JSEN.2022.3147775},
    journal = {IEEE Sensors Journal},
    month = {March},
    pages = {1--1},
    title = {Portable {{Diamond NV Magnetometer Head Integrated With}} 520 Nm {{Diode Laser}}},
    volume = {22},
    year = {2022}
}

@article{deguchiCompactPortableQuantum2023,
    author = {Deguchi, Hiroshige and Hayashi, Tsukasa and Saito, Hiroya and Nishibayashi, Yoshiki and Teramoto, Minori and Fujiwara, Masanori and Morishita, Hiroki and Mizuochi, Norikazu and Tatsumi, Natsuo},
    doi = {10.35848/1882-0786/acd836},
    issn = {1882-0786},
    journal = {Applied Physics Express},
    langid = {english},
    month = {June},
    number = {6},
    pages = {062004},
    publisher = {IOP Publishing},
    title = {Compact and Portable Quantum Sensor Module Using Diamond {{NV}} Centers},
    urldate = {2025-12-20},
    volume = {16},
    year = {2023}
}

@article{pogorzelskiCompactFullyIntegrated2024,
    author = {Pogorzelski, Jens and Horsthemke, Ludwig and Homrighausen, Jonas and Stiegek{\"o}tter, Dennis and Gregor, Markus and Gl{\"o}sek{\"o}tter, Peter and Pogorzelski, Jens and Horsthemke, Ludwig and Homrighausen, Jonas and Stiegek{\"o}tter, Dennis and Gregor, Markus and Gl{\"o}sek{\"o}tter, Peter},
    copyright = {http://creativecommons.org/licenses/by/3.0/},
    doi = {10.3390/s24030743},
    issn = {1424-8220},
    journal = {Sensors},
    langid = {english},
    month = {January},
    number = {3},
    publisher = {publisher},
    title = {Compact and {{Fully Integrated LED Quantum Sensor Based}} on {{NV Centers}} in {{Diamond}}},
    urldate = {2025-12-20},
    volume = {24},
    year = {2024}
}

@article{kimCMOSintegratedQuantumSensor2019,
    author = {Kim, Donggyu and Ibrahim, Mohamed I. and Foy, Christopher and Trusheim, Matthew E. and Han, Ruonan and Englund, Dirk R.},
    copyright = {2019 The Author(s), under exclusive licence to Springer Nature Limited},
    doi = {10.1038/s41928-019-0275-5},
    issn = {2520-1131},
    journal = {Nature Electronics},
    langid = {english},
    month = {July},
    number = {7},
    pages = {284--289},
    publisher = {Nature Publishing Group},
    title = {A {{CMOS-integrated}} Quantum Sensor Based on Nitrogen--Vacancy Centres},
    urldate = {2025-12-20},
    volume = {2},
    year = {2019}
}

@article{ibrahimHighScalabilityCMOSQuantum2021,
    author = {Ibrahim, Mohamed I. and Foy, Christopher and Englund, Dirk R. and Han, Ruonan},
    copyright = {https://ieeexplore.ieee.org/Xplorehelp/downloads/license-information/IEEE.html},
    doi = {10.1109/JSSC.2020.3027056},
    issn = {0018-9200, 1558-173X},
    journal = {IEEE Journal of Solid-State Circuits},
    langid = {english},
    month = {March},
    number = {3},
    pages = {1001--1014},
    title = {High-{{Scalability CMOS Quantum Magnetometer With Spin-State Excitation}} and {{Detection}} of {{Diamond Color Centers}}},
    urldate = {2025-11-04},
    volume = {56},
    year = {2021}
}

@article{levinePrinciplesTechniquesQuantum2019,
    author = {Levine, Edlyn V. and Turner, Matthew J. and Kehayias, Pauli and Hart, Connor A. and Langellier, Nicholas and Trubko, Raisa and Glenn, David R. and Fu, Roger R. and Walsworth, Ronald L.},
    copyright = {http://creativecommons.org/licenses/by/4.0},
    doi = {10.1515/nanoph-2019-0209},
    issn = {2192-8614, 2192-8606},
    journal = {Nanophotonics},
    langid = {english},
    month = {November},
    number = {11},
    pages = {1945--1973},
    title = {Principles and Techniques of the Quantum Diamond Microscope},
    urldate = {2025-11-04},
    volume = {8},
    year = {2019}
}

@misc{yuNoninvasiveMagnetocardiographyLiving2024,
    archiveprefix = {arXiv},
    author = {Yu, Ziyun and Xie, Yijin and Jin, Guodong and Zhu, Yunbin and Zhang, Qi and Shi, Fazhan and Wan, Fang-yan and Luo, Hongmei and Tang, Ai-hui and Rong, Xing},
    doi = {10.48550/arXiv.2405.02376},
    eprint = {2405.02376},
    langid = {english},
    month = {May},
    number = {arXiv:2405.02376},
    primaryclass = {physics},
    publisher = {arXiv},
    title = {Non-Invasive Magnetocardiography of Living Rat Based on Diamond Quantum Sensor},
    urldate = {2025-12-10},
    year = {2024}
}

@article{chenImmunomagneticMicroscopyTumor2022,
    author = {Chen, Sanyou and Li, Wanhe and Zheng, Xiaohu and Yu, Pei and Wang, Pengfei and Sun, Ziting and Xu, Yao and Jiao, Defeng and Ye, Xiangyu and Cai, Mingcheng and Shen, Mengze and Wang, Mengqi and Zhang, Qi and Kong, Fei and Wang, Ya and He, Jie and Wei, Haiming and Shi, Fazhan and Du, Jiangfeng},
    doi = {10.1073/pnas.2118876119},
    journal = {Proceedings of the National Academy of Sciences},
    month = {February},
    number = {5},
    pages = {e2118876119},
    publisher = {Proceedings of the National Academy of Sciences},
    title = {Immunomagnetic Microscopy of Tumor Tissues Using Quantum Sensors in Diamond},
    urldate = {2025-12-10},
    volume = {119},
    year = {2022}
}

@article{dohertyTheoryGroundstateSpin2012,
    author = {Doherty, M. W. and Dolde, F. and Fedder, H. and Jelezko, F. and Wrachtrup, J. and Manson, N. B. and Hollenberg, L. C. L.},
    doi = {10.1103/PhysRevB.85.205203},
    journal = {Physical Review B},
    month = {May},
    number = {20},
    pages = {205203},
    publisher = {American Physical Society},
    title = {Theory of the Ground-State Spin of the {{NV}}\$\textbraceleft\textbraceright\textasciicircum\textbraceleft\textbackslash ensuremath\textbraceleft -\textbraceright\textbraceright\$ Center in Diamond},
    urldate = {2025-12-20},
    volume = {85},
    year = {2012}
}

@article{jelezkoSingleDefectCentres2006,
    author = {Jelezko, F. and Wrachtrup, J.},
    copyright = {Copyright \copyright{} 2006 WILEY-VCH Verlag GmbH \& Co. KGaA, Weinheim},
    doi = {10.1002/pssa.200671403},
    issn = {1862-6319},
    journal = {physica status solidi (a)},
    langid = {english},
    number = {13},
    pages = {3207--3225},
    shorttitle = {Single Defect Centres in Diamond},
    title = {Single Defect Centres in Diamond: {{A}} Review},
    urldate = {2025-12-21},
    volume = {203},
    year = {2006}
}

@article{jelezkoObservationCoherentOscillations2004,
    author = {Jelezko, F. and Gaebel, T. and Popa, I. and Gruber, A. and Wrachtrup, J.},
    doi = {10.1103/PhysRevLett.92.076401},
    journal = {Physical Review Letters},
    month = {February},
    number = {7},
    pages = {076401},
    publisher = {American Physical Society},
    title = {Observation of {{Coherent Oscillations}} in a {{Single Electron Spin}}},
    urldate = {2025-12-21},
    volume = {92},
    year = {2004}
}

@article{taylorHighsensitivityDiamondMagnetometer2008,
    author = {Taylor, J. M. and Cappellaro, P. and Childress, L. and Jiang, L. and Budker, D. and Hemmer, P. R. and Yacoby, A. and Walsworth, R. and Lukin, M. D.},
    copyright = {2008 Springer Nature Limited},
    doi = {10.1038/nphys1075},
    issn = {1745-2481},
    journal = {Nature Physics},
    langid = {english},
    month = {October},
    number = {10},
    pages = {810--816},
    publisher = {Nature Publishing Group},
    title = {High-Sensitivity Diamond Magnetometer with Nanoscale Resolution},
    urldate = {2025-12-21},
    volume = {4},
    year = {2008}
}

@misc{barrySensitivityOptimizationNVDiamond2019,
    author = {Barry, John F. and Schloss, Jennifer M. and Bauch, Erik and Turner, Matthew J. and Hart, Connor A. and Pham, Linh M. and Walsworth, Ronald L.},
    doi = {10.1103/RevModPhys.92.015004},
    howpublished = {https://arxiv.org/abs/1903.08176v2},
    journal = {arXiv.org},
    langid = {english},
    month = {March},
    title = {Sensitivity {{Optimization}} for {{NV-Diamond Magnetometry}}},
    urldate = {2025-11-12},
    year = {2019}
}

@article{zickusFluorescenceLifetimeImaging2020,
    author = {Zickus, Vytautas and Wu, Ming-Lo and Morimoto, Kazuhiro and Kapitany, Valentin and Fatima, Areeba and Turpin, Alex and Insall, Robert and Whitelaw, Jamie and Machesky, Laura and Bruschini, Claudio and Faccio, Daniele and Charbon, Edoardo},
    copyright = {2020 The Author(s)},
    doi = {10.1038/s41598-020-77737-0},
    issn = {2045-2322},
    journal = {Scientific Reports},
    langid = {english},
    month = {December},
    number = {1},
    pages = {20986},
    publisher = {Nature Publishing Group},
    title = {Fluorescence Lifetime Imaging with a Megapixel {{SPAD}} Camera and Neural Network Lifetime Estimation},
    urldate = {2025-12-21},
    volume = {10},
    year = {2020}
}

@article{morimotoMegapixelTimegatedSPAD2020,
    author = {Morimoto, Kazuhiro and Ardelean, Andrei and Wu, Ming-Lo and Ulku, Arin Can and Antolovic, Ivan Michel and Bruschini, Claudio and Charbon, Edoardo},
    copyright = {\copyright{} 2020 Optical Society of America},
    doi = {10.1364/OPTICA.386574},
    issn = {2334-2536},
    journal = {Optica},
    langid = {english},
    month = {April},
    number = {4},
    pages = {346--354},
    publisher = {Optica Publishing Group},
    title = {Megapixel Time-Gated {{SPAD}} Image Sensor for {{2D}} and {{3D}} Imaging Applications},
    urldate = {2025-12-21},
    volume = {7},
    year = {2020}
}

@article{henderson$192times128$TimeCorrelated2019,
    author = {Henderson, Robert K. and Johnston, Nick and Mattioli Della Rocca, Francescopaolo and Chen, Haochang and {Day-Uei Li}, David and Hungerford, Graham and Hirsch, Richard and Mcloskey, David and Yip, Philip and Birch, David J. S.},
    copyright = {https://ieeexplore.ieee.org/Xplorehelp/downloads/license-information/IEEE.html},
    doi = {10.1109/JSSC.2019.2905163},
    issn = {0018-9200, 1558-173X},
    journal = {IEEE Journal of Solid-State Circuits},
    langid = {english},
    month = {July},
    number = {7},
    pages = {1907--1916},
    title = {A \$192\textbackslash times128\$ {{Time Correlated SPAD Image Sensor}} in 40-Nm {{CMOS Technology}}},
    urldate = {2025-11-04},
    volume = {54},
    year = {2019}
}

@article{hadfieldSinglephotonDetectorsOptical2009,
    author = {Hadfield, Robert H.},
    copyright = {2009 Springer Nature Limited},
    doi = {10.1038/nphoton.2009.230},
    issn = {1749-4893},
    journal = {Nature Photonics},
    langid = {english},
    month = {December},
    number = {12},
    pages = {696--705},
    publisher = {Nature Publishing Group},
    title = {Single-Photon Detectors for Optical Quantum Information Applications},
    urldate = {2025-12-21},
    volume = {3},
    year = {2009}
}

@article{zhangAdvancesInGaAsInP2015,
    author = {Zhang, Jun and Itzler, Mark A. and Zbinden, Hugo and Pan, Jian-Wei},
    copyright = {2015 The Author(s)},
    doi = {10.1038/lsa.2015.59},
    issn = {2047-7538},
    journal = {Light: Science \& Applications},
    langid = {english},
    month = {May},
    number = {5},
    pages = {e286-e286},
    publisher = {Nature Publishing Group},
    title = {Advances in {{InGaAs}}/{{InP}} Single-Photon Detector Systems for Quantum Communication},
    urldate = {2025-12-21},
    volume = {4},
    year = {2015}
}

@article{bruschiniSinglephotonAvalancheDiode2019,
    author = {Bruschini, Claudio and Homulle, Harald and Antolovic, Ivan Michel and Burri, Samuel and Charbon, Edoardo},
    copyright = {2019 The Author(s)},
    doi = {10.1038/s41377-019-0191-5},
    issn = {2047-7538},
    journal = {Light: Science \& Applications},
    langid = {english},
    month = {September},
    number = {1},
    pages = {87},
    publisher = {Nature Publishing Group},
    shorttitle = {Single-Photon Avalanche Diode Imagers in Biophotonics},
    title = {Single-Photon Avalanche Diode Imagers in Biophotonics: Review and Outlook},
    urldate = {2025-12-21},
    volume = {8},
    year = {2019}
}

@article{zappaPrinciplesFeaturesSinglephoton2007,
    author = {Zappa, F. and Tisa, S. and Tosi, A. and Cova, S.},
    copyright = {https://www.elsevier.com/tdm/userlicense/1.0/},
    doi = {10.1016/j.sna.2007.06.021},
    issn = {09244247},
    journal = {Sensors and Actuators A: Physical},
    langid = {english},
    month = {October},
    number = {1},
    pages = {103--112},
    title = {Principles and Features of Single-Photon Avalanche Diode Arrays},
    urldate = {2025-11-04},
    volume = {140},
    year = {2007}
}

@article{leeHighPerformanceBackIlluminatedThreeDimensional2018,
    author = {Lee, Myung-Jae and Ximenes, Augusto Ronchini and Padmanabhan, Preethi and Wang, Tzu-Jui and Huang, Kuo-Chin and Yamashita, Yuichiro and Yaung, Dun-Nian and Charbon, Edoardo},
    copyright = {https://ieeexplore.ieee.org/Xplorehelp/downloads/license-information/OAPA.html},
    doi = {10.1109/JSTQE.2018.2827669},
    issn = {1077-260X, 1558-4542},
    journal = {IEEE Journal of Selected Topics in Quantum Electronics},
    langid = {english},
    month = {November},
    number = {6},
    pages = {1--9},
    title = {High-{{Performance Back-Illuminated Three-Dimensional Stacked Single-Photon Avalanche Diode Implemented}} in 45-Nm {{CMOS Technology}}},
    urldate = {2025-11-04},
    volume = {24},
    year = {2018}
}

@inproceedings{pellegriniIndustrialisedSPAD402017,
    address = {San Francisco, CA, USA},
    author = {Pellegrini, S. and Rae, B. and Pingault, A. and Golanski, D. and Jouan, S. and Lapeyre, C. and Mamdy, B.},
    booktitle = {2017 {{IEEE International Electron Devices Meeting}} ({{IEDM}})},
    doi = {10.1109/IEDM.2017.8268404},
    isbn = {978-1-5386-3559-9},
    langid = {english},
    month = {December},
    pages = {16.5.1-16.5.4},
    publisher = {IEEE},
    title = {Industrialised {{SPAD}} in 40 Nm Technology},
    urldate = {2025-11-04},
    year = {2017}
}

@article{alAbbasGlobalSharedWellSPAD2017,
    author = {Al Abbas, Tarek and Dutton, Neale and Almer, Oscar and Mattioli Della Rocca, Francesco and Pellegrini, Sara and Rae, Bruce R. and Golanski, D. and Henderson, Robert},
    journal = {International Image Sensor Workshop (IISW 2017)},
    langid = {english},
    month = {June},
    pages = {},
    title = {8.25 μm Pitch 66\% Fill Factor Global Shared Well SPAD Image Sensor in 40 nm CMOS FSI Technology},
    url = {https://www.research.ed.ac.uk/en/publications/825%CE%BCm-pitch-66-fill-factor-global-shared-well-spad-image-sensor-i},
    urldate = {2026-02-18},
    year = {2017}
}

@inproceedings{morimoto32Megapixel3DStacked2021,
    author = {Morimoto, K. and Iwata, J. and Shinohara, M. and Sekine, H. and Abdelghafar, A. and Tsuchiya, H. and Kuroda, Y. and Tojima, K. and Endo, W. and Maehashi, Y. and Ota, Y. and Sasago, T. and Maekawa, S. and Hikosaka, S. and Kanou, T. and Kato, A. and Tezuka, T. and Yoshizaki, S. and Ogawa, T. and Uehira, K. and Ehara, A. and Inui, F. and Matsuno, Y. and Sakurai, K. and Ichikawa, T.},
    booktitle = {2021 {{IEEE International Electron Devices Meeting}} ({{IEDM}})},
    doi = {10.1109/IEDM19574.2021.9720605},
    issn = {2156-017X},
    month = {December},
    pages = {20.2.1-20.2.4},
    title = {3.2 {{Megapixel 3D-Stacked Charge Focusing SPAD}} for {{Low-Light Imaging}} and {{Depth Sensing}}},
    urldate = {2025-12-21},
    year = {2021}
}

@article{deAlbuquerqueIntegrationSPAD2018,
    author = {de Albuquerque, T. Chaves and Calmon, F. and Clerc, R. and Pittet, P. and Benhammou, Y. and Golanski, D. and Jouan, S. and Rideau, D. and Cathelin, A.},
    doi = {10.1109/ESSDERC.2018.8486852},
    journal = {Proceedings of the 48th European Solid-State Device Research Conference (ESSDERC 2018)},
    langid = {english},
    pages = {82--85},
    title = {Integration of SPAD in 28 nm FDSOI CMOS Technology},
    year = {2018}
}

@article{issartelArchitectureOptimizationSPAD2022,
    author = {Issartel, D. and Gao, S. and Pittet, P. and Cellier, R. and Golanski, D. and Cathelin, A. and Calmon, F.},
    doi = {10.1016/j.sse.2022.108297},
    issn = {0038-1101},
    journal = {Solid-State Electronics},
    langid = {english},
    pages = {108297},
    title = {Architecture Optimization of SPAD Integrated in 28 nm FD-SOI CMOS Technology to Reduce the DCR},
    volume = {191},
    year = {2022}
}

@article{gaoCorrelationsDCRPDP2023,
    author = {Gao, Shaochen and Issartel, Dylan and Dolatpoor Lakeh, Mohammadreza and Mandorlo, Fabien and Orobtchouk, R{\'e}gis and Kammerer, Jean-Baptiste and Cathelin, Andreia and Golanski, Dominique and Uhring, Wilfried and Calmon, Francis},
    journal = {International Image Sensor Workshop (IISW 2023)},
    langid = {english},
    month = {May},
    note = {Poster presented at IISW 2023, Crieff, United Kingdom, 21--25 May 2023},
    title = {Correlations between DCR and PDP of SPAD Integrated in a 28 nm FD-SOI CMOS Technology},
    url = {https://hal.science/hal-03977945},
    urldate = {2026-02-18},
    year = {2023}
}

@article{covaAvalanchePhotodiodesQuenching1996,
    author = {Cova, S. and Ghioni, M. and Lacaita, A. and Samori, C. and Zappa, F.},
    copyright = {\copyright{} 1996 Optical Society of America},
    doi = {10.1364/AO.35.001956},
    issn = {2155-3165},
    journal = {Applied Optics},
    langid = {english},
    month = {April},
    number = {12},
    pages = {1956--1976},
    publisher = {Optica Publishing Group},
    title = {Avalanche Photodiodes and Quenching Circuits for Single-Photon Detection},
    urldate = {2025-12-21},
    volume = {35},
    year = {1996}
}

@article{dautetPhotonCountingTechniques1993,
    author = {Dautet, Henri and Deschamps, Pierre and Dion, Bruno and MacGregor, Andrew D. and MacSween, Darleene and McIntyre, Robert J. and Trottier, Claude and Webb, Paul P.},
    copyright = {\copyright{} 1993 Optical Society of America},
    doi = {10.1364/AO.32.003894},
    issn = {2155-3165},
    journal = {Applied Optics},
    langid = {english},
    month = {July},
    number = {21},
    pages = {3894--3900},
    publisher = {Optica Publishing Group},
    title = {Photon Counting Techniques with Silicon Avalanche Photodiodes},
    urldate = {2025-12-21},
    volume = {32},
    year = {1993}
}

@article{charbonSinglephotonImagingComplementary2014,
    author = {Charbon, E.},
    doi = {10.1098/rsta.2013.0100},
    issn = {1364-503X},
    journal = {Philosophical Transactions of the Royal Society A: Mathematical, Physical and Engineering Sciences},
    month = {March},
    number = {2012},
    pages = {20130100},
    title = {Single-Photon Imaging in Complementary Metal Oxide Semiconductor Processes},
    urldate = {2025-12-21},
    volume = {372},
    year = {2014}
}

@article{stipcevicCharacterizationNovelAvalanche2010,
    author = {Stip{\v c}evi{\'c}, M. and Skenderovi{\'c}, H. and Gracin, D.},
    copyright = {https://doi.org/10.1364/OA\_License\_v1\#VOR-OA},
    doi = {10.1364/OE.18.017448},
    issn = {1094-4087},
    journal = {Optics Express},
    langid = {english},
    month = {August},
    number = {16},
    pages = {17448},
    title = {Characterization of a Novel Avalanche Photodiode for Single Photon Detection in {{VIS-NIR}} Range},
    urldate = {2025-11-04},
    volume = {18},
    year = {2010}
}

@article{ghioniCompactActiveQuenching1996,
    author = {Ghioni, M. and Cova, S. and Zappa, F. and Samori, C.},
    doi = {10.1063/1.1147156},
    issn = {0034-6748},
    journal = {Review of Scientific Instruments},
    month = {October},
    number = {10},
    pages = {3440--3448},
    title = {Compact Active Quenching Circuit for Fast Photon Counting with Avalanche Photodiodes},
    urldate = {2025-12-21},
    volume = {67},
    year = {1996}
}

@inproceedings{inproceedings,
    author = {Ibrahim, Isack and Foy, Christopher and Kim, Donggyu and Englund, Dirk and Han, Ruonan},
    doi = {10.1109/VLSIC.2018.8502329},
    month = {06},
    pages = {249-250},
    title = {Room-Temperature Quantum Sensing in CMOS: On-Chip Detection of Electronic Spin States in Diamond Color Centers for Magnetometry},
    year = {2018}
}

@article{dolatpoorlakehUltrafastActiveQuenching2021,
    author = {Dolatpoor Lakeh, Mohammadreza and Kammerer, Jean-Baptiste and Agu{\'e}nounon, Enagnon and Issartel, Dylan and Schell, Jean-Baptiste and Rink, Sven and Cathelin, Andreia and Calmon, Francis and Uhring, Wilfried},
    doi = {10.3390/s21124014},
    issn = {1424-8220},
    journal = {Sensors},
    langid = {english},
    month = {June},
    number = {12},
    pages = {4014},
    title = {An {{Ultrafast Active Quenching Active Reset Circuit}} with 50\% {{SPAD Afterpulsing Reduction}} in a 28 Nm {{FD-SOI CMOS Technology Using Body Biasing Technique}}},
    urldate = {2025-11-04},
    volume = {21},
    year = {2021}
}

@article{liFibercoupledScanningMagnetometer2023,
    archiveprefix = {arXiv},
    author = {Li, Yufan and Gerritsma, Fabian A. and Kurdi, Samer and Codreanu, Nina and Gr{\"o}blacher, Simon and Hanson, Ronald and Norte, Richard and van der Sar, Toeno},
    doi = {10.1021/acsphotonics.3c00259},
    eprint = {2302.12536},
    issn = {2330-4022, 2330-4022},
    journal = {ACS Photonics},
    month = {June},
    number = {6},
    pages = {1859--1865},
    primaryclass = {cond-mat},
    title = {A {{Fiber-coupled Scanning Magnetometer}} with {{Nitrogen-Vacancy Spins}} in a {{Diamond Nanobeam}}},
    urldate = {2026-02-06},
    volume = {10},
    year = {2023}
}

@article{araiMillimetrescaleMagnetocardiographyLiving2022,
    author = {Arai, Keigo and Kuwahata, Akihiro and Nishitani, Daisuke and Fujisaki, Ikuya and Matsuki, Ryoma and Nishio, Yuki and Xin, Zonghao and Cao, Xinyu and Hatano, Yuji and Onoda, Shinobu and Shinei, Chikara and Miyakawa, Masashi and Taniguchi, Takashi and Yamazaki, Masatoshi and Teraji, Tokuyuki and Ohshima, Takeshi and Hatano, Mutsuko and Sekino, Masaki and Iwasaki, Takayuki},
    doi = {10.1038/s42005-022-00978-0},
    issn = {2399-3650},
    journal = {Communications Physics},
    langid = {english},
    month = {August},
    number = {1},
    pages = {200},
    title = {Millimetre-Scale Magnetocardiography of Living Rats with Thoracotomy},
    urldate = {2025-12-10},
    volume = {5},
    year = {2022}
}

@book{fishburnFundamentalsCMOSSinglephoton2012,
    address = {S.l.},
    author = {Fishburn, Matthew W.},
    isbn = {978-94-91030-29-1},
    langid = {english},
    publisher = {s.n.},
    title = {Fundamentals of {{CMOS}} Single-Photon Avalanche Diodes},
    year = {2012}
}

@article{takaiSinglePhotonAvalancheDiode2016,
    author = {Takai, Isamu and Matsubara, Hiroyuki and Soga, Mineki and Ohta, Mitsuhiko and Ogawa, Masaru and Yamashita, Tatsuya and Takai, Isamu and Matsubara, Hiroyuki and Soga, Mineki and Ohta, Mitsuhiko and Ogawa, Masaru and Yamashita, Tatsuya},
    copyright = {http://creativecommons.org/licenses/by/3.0/},
    doi = {10.3390/s16040459},
    issn = {1424-8220},
    journal = {Sensors},
    langid = {english},
    month = {March},
    number = {4},
    publisher = {publisher},
    title = {Single-{{Photon Avalanche Diode}} with {{Enhanced NIR-Sensitivity}} for {{Automotive LIDAR Systems}}},
    urldate = {2025-12-21},
    volume = {16},
    year = {2016}
}

@article{indexgel,
    author = {Sankawa, Izumi and Satake, Toshiaki and Kashima, Norio and Nagasawa, Shinji},
    doi = {10.1002/ecja.4410690111},
    journal = {Electronics and Communications in Japan Part I-communications - ELECTRON COMMUN JPN I},
    month = {01},
    pages = {94-102},
    title = {Methods for reducing the fresnel reflection in an optical-fiber connector with index matching material},
    volume = {69},
    year = {1986}
}

@misc{qnamiFundamentalsMagneticField2020,
    author = {Qnami},
    langid = {english},
    month = {December},
    publisher = {Qnami},
    title = {Fundamentals of Magnetic Field Measurement with {{NV}} Centers in Diamond (2020)},
    urldate = {2025-12-21},
    year = {2020}
}

@article{dreauAvoidingPowerBroadening2011,
    archiveprefix = {arXiv},
    author = {Dr{\'e}au, A. and Lesik, M. and Rondin, L. and Spinicelli, P. and Arcizet, O. and Roch, J.-F. and Jacques, V.},
    doi = {10.1103/PhysRevB.84.195204},
    eprint = {1108.0178},
    issn = {1098-0121, 1550-235X},
    journal = {Physical Review B},
    month = {November},
    number = {19},
    pages = {195204},
    primaryclass = {cond-mat},
    title = {Avoiding Power Broadening in Optically Detected Magnetic Resonance of Single {{NV}} Defects for Enhanced {{DC-magnetic}} Field Sensitivity},
    urldate = {2025-11-04},
    volume = {84},
    year = {2011}
}

@article{bauchUltralongDephasingTimes2018,
    archiveprefix = {arXiv},
    author = {Bauch, Erik and Hart, Connor A. and Schloss, Jennifer M. and Turner, Matthew J. and Barry, John F. and Kehayias, Pauli and Singh, Swati and Walsworth, Ronald L.},
    doi = {10.1103/PhysRevX.8.031025},
    eprint = {1801.03793},
    issn = {2160-3308},
    journal = {Physical Review X},
    month = {July},
    number = {3},
    pages = {031025},
    primaryclass = {quant-ph},
    title = {Ultralong {{Dephasing Times}} in {{Solid-State Spin Ensembles}} via {{Quantum Control}}},
    urldate = {2025-12-21},
    volume = {8},
    year = {2018}
}

@article{bauchDecoherenceDipolarSpin2020,
    archiveprefix = {arXiv},
    author = {Bauch, Erik and Singh, Swati and Lee, Junghyun and Hart, Connor A. and Schloss, Jennifer M. and Turner, Matthew J. and Barry, John F. and Pham, Linh and {Bar-Gill}, Nir and Yelin, Susanne F. and Walsworth, Ronald L.},
    doi = {10.1103/PhysRevB.102.134210},
    eprint = {1904.08763},
    issn = {2469-9950, 2469-9969},
    journal = {Physical Review B},
    month = {October},
    number = {13},
    pages = {134210},
    primaryclass = {quant-ph},
    title = {Decoherence of Dipolar Spin Ensembles in Diamond},
    urldate = {2025-11-04},
    volume = {102},
    year = {2020}
}

@article{weiSuperparamagneticIronOxide2021,
    author = {Wei, Hao and Hu, Yangnan and Wang, Junguo and Gao, Xia and Qian, Xiaoyun and Tang, Mingliang},
    copyright = {http://creativecommons.org/licenses/by-nc/3.0/},
    doi = {10.2147/IJN.S321984},
    issn = {1178-2013},
    journal = {International Journal of Nanomedicine},
    langid = {english},
    month = {August},
    pages = {6097--6113},
    shorttitle = {Superparamagnetic {{Iron Oxide Nanoparticles}}},
    title = {Superparamagnetic {{Iron Oxide Nanoparticles}}: {{Cytotoxicity}}, {{Metabolism}}, and {{Cellular Behavior}} in {{Biomedicine Applications}}},
    urldate = {2025-11-04},
    volume = {Volume 16},
    year = {2021}
}

\EOD

\end{document}